\documentclass[manuscript]{acmart}

\usepackage{graphicx}
\usepackage{dcolumn}
\usepackage{bm}
\usepackage{tabularx}
\usepackage{float}
\usepackage{braket}
\usepackage{hyperref}
\usepackage{algorithm}
 \usepackage{algpseudocode}
 \usepackage{algorithmicx}
 \algdef{SE}[DOWHILE]{Do}{doWhile}{\algorithmicdo}[1]{\algorithmicwhile\ #1}%

\algrenewcommand\algorithmicrequire{\textbf{Input:}}
\algrenewcommand\algorithmicensure{\textbf{Output:}}
\algnewcommand{\algorithmicor}{\textbf{ or }}

\algnewcommand{\OR}{\algorithmicor}
\usepackage{xcolor}
\newcommand{\tnl}{\text{{TNL }}}

\usepackage[justification=raggedright,singlelinecheck=false]{caption}
\usepackage{subcaption}

\begin{document}


\title{Iterative Interpolation Schedules for Quantum Approximate Optimization Algorithm}


\author{Anuj Apte}
\authornote{Equal contribution.}

\author{Shree Hari Sureshbabu}
\authornote{Equal contribution. Correspondence should be addressed to \texttt{shreehari.sureshbabu@jpmchase.com}}

\author{Ruslan Shaydulin}
\author{Sami Boulebnane}
\author{Zichang He}
\author{Dylan Herman}
\author{James Sud}
\author{Marco Pistoia}

\affiliation{%
  \institution{Global Technology Applied Research, JPMorganChase}
  \city{New York}
  \state{New York}
  \country{USA}
}

\begin{abstract}
Quantum Approximate Optimization Algorithm (QAOA) is a promising quantum heuristic with empirical evidence of speedup over classical state-of-the-art for some problems. QAOA uses a parameterized circuit with $p$ layers, where higher $p$ yields better solutions, but requires optimizing $2p$ independent parameters, which is challenging at large $p$. We present an iterative interpolation method that exploits the smoothness of optimal parameter schedules by expressing them in a basis of orthogonal functions, generalizing the work of Zhou et al. By optimizing a small number of basis coefficients and iteratively increasing both circuit depth and coefficient count until convergence, our method constructs high-quality schedules for large $p$. We provide theoretical justification using Jackson's theorem and Lipschitz continuity to bound the required number of basis coefficients for a given accuracy. Our approach achieves better performance with fewer optimization steps than existing methods across three benchmark problems: the Sherrington-Kirkpatrick (SK) model, portfolio optimization, and Low Autocorrelation Binary Sequences (LABS). For the largest LABS instance, we achieve near-optimal merit factors with schedules exceeding 1000 layers, an order of magnitude beyond previous methods. Additionally, we observe that a mild growth in QAOA depth suffices to solve the SK model exactly, a result of independent theoretical interest.
\end{abstract}

\maketitle



\section{Introduction}\label{sec:intro}

The quantum approximate optimization algorithm (QAOA) is a promising heuristic for combinatorial optimization~\cite{Hogg2000,Hogg2000search,farhi2014quantum}. Originally proposed by Hogg in 2000~\cite{Hogg2000search,Hogg2000}, it has attracted increasing attention in the recent years. The low hardware requirements of QAOA enable small-scale experiments on today's devices~\cite{Shaydulin2023npgeq,Pelofske2023,Pelofske2024,2409.12104} and make QAOA an appealing candidate algorithm for early fault-tolerant quantum computers~\cite{ICCAD_qaoapara}. Furthermore, numerical and theoretical evidence has emerged indicating that QAOA offers a polynomial~\cite{boulebnane2022solving,shaydulin2023evidence,2411.17442} and, in some restricted settings, perhaps even exponential~\cite{2411.04979} speedup over state-of-the-art classical solvers.

QAOA solves optimization problems using a quantum circuit composed of a sequence of alternating, parameterized quantum evolutions, referred to as \textit{layers}, with the number of layers denoted by $p$ and each layer containing two operators: a ``phase'' operator encoding the objective function and a non-commuting ``mixing'' operator inducing nontrivial dynamics. These alternating operators aim to transform an initial, easy-to-prepare quantum state into a state that encodes the solution to a given combinatorial optimization problem. We will refer to the combined $2p$ parameters as a \textit{schedule}. 
Analytically optimal or otherwise well-performing instance-independent parameters exist for many problems~\cite{sureshbabu2024parameter,shaydulin2023evidence,boulebnane2022solving,basso2021quantum,farhi2022quantum,he2025non,hao2024end}. However, they are only known for small depth $p \lesssim 40$. Consequently, executing QAOA with many layers requires numerically optimizing $2p$ free parameters, which can become exceedingly challenging or even infeasible as $p$ grows.

\begin{figure}[b]
    \centering
    \includegraphics[width=0.6\textwidth]{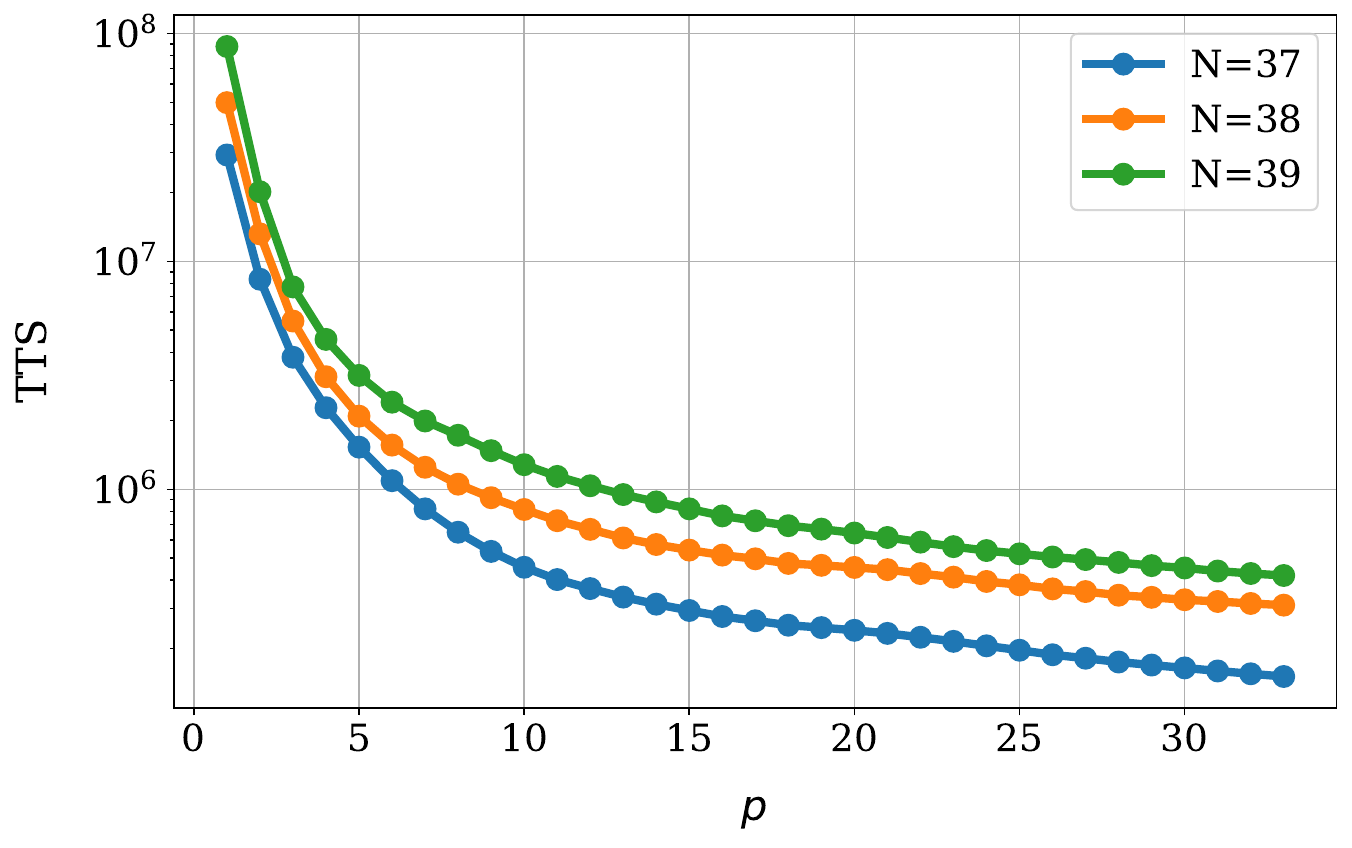}
    \caption{\label{fig:tts_motivation}
    As $p$ grows, QAOA TTS decreases despite increasing circuit depth. Here TTS is the total number of QAOA layers that must be executed in expectation to solve the LABS problem exactly. Data reproduced from Ref.~\cite{shaydulin2023evidence}.}
\end{figure}

The need for high $p$ in QAOA is motivated by the observation that while evidence exists for QAOA offering a speedup with small constant $p$~\cite{boulebnane2022solving,shaydulin2023evidence,2411.17442,2411.04979}, increasing $p$ improves the performance. For example, Fig.~\ref{fig:tts_motivation} reinterprets the data from Ref.~\cite{shaydulin2023evidence} to show that QAOA time to solution (TTS) improves as $p$ grows despite the overhead of having to execute a deeper circuit. Specifically, in Fig.~\ref{fig:tts_motivation}, TTS is equal to the number of times QAOA needs to be executed in expectation to see an optimal solution (equal to the inverse of the overlap between QAOA state and the ground state of the problem Hamiltonian) times the number of layers $p$. In addition to improving TTS for given fixed problem size $N$, increasing $p$ has been shown to increase the degree of the asymptotic quantum speedup offered by QAOA~\cite{shaydulin2023evidence,boulebnane2022solving}, though how large $p$ can get before increasing it further does not improve asymptotic behavior may depend on the problem and schedule choice.

Multiple strategies have been proposed to overcome the challenge of optimizing QAOA parameters at high depth. Zhou et al.~\cite{zhou2020quantum} proposed a strategy that uses the parameters optimized for QAOA with $p-1$ layers as the initial point for $p$ layers. Ref.~\cite{zhou2020quantum} proposes multiple strategies, but the best performing one views the QAOA schedule as a time series and performs a sine (cosine) transform of the parameters corresponding to phase (mixing) operator. Transforming parameters for depth $p-1$ yields $p-1$ coefficients; the interpolation is then performed by appending zeros to the vector of coefficients and transforming back into the original parameter space to obtain schedule for depth $p$. While this technique has been shown to perform well for multiple problems~\cite{farhi2022quantum,shaydulin2023evidence}, it has only been evaluated for modest QAOA depth. Other notable approaches include using linear ramps as initial QAOA schedules~\cite{Sack2021, kremenetski2021quantum, kremenetski2023quantum} or using an efficiently computable homogeneous proxy to optimize parameters on a classical computer~\cite{Sud2024}.

In this work, we introduce a general approach to QAOA parameter optimization at high depth that generalizes the technique of Zhou et al.~\cite{zhou2020quantum}. Our central observation is two-fold. First, we allow for transforms beyond the discrete sine / cosine considered in Ref.~\cite{zhou2020quantum}. Second, we adaptively choose the number of coefficients to optimize. Remarkably, these simple modifications enable our \emph{iterative interpolation} technique to perform dramatically better, enabling the study of QAOA with optimized parameters at very large depth. Specifically, we obtain QAOA depth sufficient to solve to near-optimality the Low Autocorrelation Binary Sequences (LABS) problem with up to 25 spins and the Sherrington-Kirkpatrick (SK) model with up to 28 spins.

We provide a theoretical foundation for our method by combining Jackson's approximation theorem with the Lipschitz continuity of QAOA expectation values to derive an explicit bound on the number of basis coefficients needed to achieve a target accuracy in the cost function. This result formally explains the empirical observation that QAOA schedules admit highly compact representations in orthogonal bases, with the required number of coefficients governed by the smoothness of the schedule and the spectral norm of the cost and mixer Hamiltonians.

Our technique makes possible the study of QAOA with hundreds of layers, allowing us to investigate the growth of QAOA depth sufficient to solve optimization problems exactly or near-optimally. We observe that the QAOA depth sufficient to find the ground state of the SK model grows mildly with problem size, which is a result of independent interest. We remark that further work is required to validate the precise scaling. In contrast to the SK model, we observe a rapid increase of depth for the LABS problem, with an exponential scaling approximately matching that of QAOA combined with quantum minimum finding reported in Ref.~\cite{shaydulin2023evidence}.

The paper is structured as follows: Section \ref{sec:background} provides the relevant background on problem formulations and prior work. Section \ref{sec:proposed_II} details our proposed method. Section \ref{sec:results} presents the numerical results on the performance of our method and on the grows of QAOA depth sufficient to solve LABS and SK exactly. Finally, Section \ref{sec:discussion} concludes with a discussion of the implications of our results.

\section{Background}\label{sec:background}


\subsection{Quantum Approximate Optimization Algorithm}
The Quantum Approximate Optimization Algorithm (QAOA) is a hybrid quantum-classical 
algorithm for 
combinatorial optimization problems \cite{Hogg2000,Hogg2000search,farhi2014quantum}. 
When applied to classical optimization problems, the algorithm operates through alternating applications of two unitary operators: evolution with the diagonal cost Hamiltonian $H_C$ which encodes the optimization objective (i.e., $H_C\ket{x}=f(x)\ket{x}, \forall x\in \{0,1\}^n$), and evolution with non-diagonal mixer Hamiltonian $H_B$ which generates
transitions between different computational basis states. 

The initial state $|s\rangle$ is chosen to be the ground state of the mixer Hamiltonian $H_B$, such that it is easy to prepare on quantum hardware. In the commonly used case where $H_B = \sum_{i=1}^N \sigma_x^i$ and $\sigma_x^i$ is a Pauli $X$ acting on $i$th qubit, this corresponds to the initial state being equal superposition over all computational basis states $|s\rangle = |+\rangle^{\otimes N}$.

The QAOA state after $p$ layers is given by:
\begin{equation}
\vert \bm\gamma, \bm\beta \rangle = \prod_{j=1}^{p} e^{-i\beta_j H_B}e^{-i\gamma_j H_C}|s\rangle~,
\end{equation}
where $\bm\gamma = (\gamma_1, ..., \gamma_p)$ and $\bm\beta = (\beta_1, ..., \beta_p)$ are $2p$ free parameters that need to be optimized or otherwise chosen. These parameters are referred to as angles, since the corresponding unitaries implement rotations in the space of quantum states.


The algorithm's performance can evaluated using either the approximation ratio (AR) or time to solution (TTS), defined as follows:
\begin{equation}
\text{AR} = \frac{\langle \bm\gamma, \bm\beta \vert H_C \vert \bm\gamma, \bm\beta \rangle}{\mathcal{C}_{\text{max}}}~, \;\;\;\;\; \text{TTS} = \frac{p}{|\langle x^* \vert \bm\gamma,\bm\beta \rangle|^2}~,
\end{equation}
where $\mathcal{C}_{\text{max}}$ represents the optimal (maximum) value of the cost function and $\ket{x^*}$ is the ground state of $H_C$.

\subsection{Low Autocorrelation Binary Sequences (LABS) problem}
Consider a sequence of binary variables $\bm s = (s_1, s_2, ..., s_N)$ of length $N$ with the auto-correlations given by $C_k(\bm s) = \sum_{i=1}^{N-k} s_i s_{i+k}$.
The goal of the low auto-correlation binary sequences (LABS) problem is to find a sequence that minimizes energy: 
\begin{align}
    \min_{\bm s \in \{-1,1\}^N} E_{\text{LABS}}(\bm s) = \min_{\bm s \in \{-1,1\}^N}\sum_{k=1}^{N-1} C_k^2~, 
\end{align}
We will also use a normalized quantity denoted merit factor MF$(\bm s)=N^2/(2E_{\text{LABS}}(\bm s))$, which remains approximately constant as $N$ grows. 

Low autocorrelation sequences are useful as modulation pulses in radar and
sonar ranging \cite{Golay1972, Pasha2000}. The best existing solvers for LABS have a runtime that scales exponentially with sequence length $N$. Optimal sequences for $N\leq 66$ have been found by branch-and-bound method \cite{Packebusch2016}. Recently, it was shown that QAOA with a fixed schedule has a runtime that scales with a smaller exponent than the best available classical methods \cite{shaydulin2023evidence}. 
A remarkable aspect that makes the LABS problem particularly interesting is that there exists a single problem instance for a given $N$. This contrasts with most other combinatorial optimization problems such as the traveling salesman problem, where instances depend on additional details like the structure of a graph. 

\subsection{Portfolio Optimization}
The central task of portfolio construction is designing a portfolio of securities which produces the greatest risk-adjusted return \cite{Markowitz1952}. In some settings (e.g. fixed-income), investors can only hold an integer number of shares of various securities. We will consider a simplified version with only binary variables indicating whether or not a given security is included in the portfolio. The risk-adjusted returns can then be expressed as a binary quadratic optimization problem:
\begin{align*}
    \min_{\mathbf{x} \in \{0,1\}^N} \quad q \mathbf{x}^T \mathbf{\Sigma} \mathbf{x} - \boldsymbol{\mu}^T \mathbf{x}\quad,\quad
    \mathrm{s.t.} \quad \mathbf{1}^T \mathbf{x} = K~,
\end{align*}
where $N$ represents the total number of available assets from which exactly $K$ must be selected. The objective function balances two competing goals: minimizing portfolio risk while maximizing expected returns. The risk-return trade-off parameter $q$ controls the relative importance of these objectives \cite{Black1992}. The $N \times N$ matrix $\mathbf{\Sigma}$ represents the covariance between asset returns, capturing their pairwise correlations and individual volatility \cite{DeMiguel2007}. The vector $\boldsymbol{\mu} \in \mathbb{R}^N$ contains the expected returns for each asset. As a binary optimization problem, it has been studied extensively with QAOA \cite{Brandhofer2022, Buonaiuto2023, Yuan_2026, He2023,sureshbabu2024parameter, Hao2022}. 

\subsection{Sherrington-Kirkpatrick (SK) model}
The Sherrington-Kirkpatrick (SK) model is a paradigmatic example of a spin glass system, representing magnetic materials with disordered interactions \cite{Sherrington1975}. The model consists of $N$ binary spin variables $s_i \in \{-1,1\}$, with all-to-all interactions between spins sampled from the standard normal distribution, $J_{ij}\sim N(0,1)$. The optimization task is to find a spin configuration $\mathbf{s}$ that minimizes the energy:
\begin{align*}
\min_{\bm s\in \{-1,1\}^N} \frac{1}{\sqrt{N}}\sum_{i<j} J_{ij} s_i s_j~,
\end{align*}
The normalization of the variance by $N$ ensures a well-defined thermodynamic limit \cite{Parisi1980}.

The SK model is particularly significant in the study of complex systems as it exhibits a rugged energy landscape and replica symmetry breaking \cite{Mezard1986}. When phrased as a decision problem, determining whether there exists a spin configuration with energy below a given threshold is NP-hard in the worst case \cite{Barahona1982}, making it an excellent benchmark for general purpose optimization algorithms \cite{HibatAllah2021, Apte2022}. In the average-case however, a message-passing algorithm of \cite{montanari2019} is able to achieve a $(1-\epsilon)$ approximation in $\mathcal{O}(C(\epsilon) n^2)$ time, where $C(\epsilon)$ is a function of the relative error $\epsilon$ and is independent of $n$. Recent studies have explored QAOA's performance on the SK model, investigating whether it can outperform classical local search algorithms \cite{Venturelli2015, Basso2022, Farhi2022}. Unlike typical optimization problems, the SK model's random nature allows for systematic study of algorithm performance by averaging over many instances drawn from the same distribution of couplings.

\subsection{Parameter Setting in QAOA} 

To overcome the challenge of direct optimization of QAOA parameters, it has been proposed to leverage the observation that optimized QAOA parameters are strikingly similar across random instances of a single problem size \cite{brandao2018fixed,Shaydulin2019,galda2021transferability, shaydulin2023parameter, sureshbabu2024parameter}, and even across instances of difference sizes \cite{brandao2018fixed,shaydulin2023evidence} and varying schedule lengths $p$ \cite{Pagano_2020, zhou2020quantum, brady2021behavior,shaydulin2023evidence}. Thus, the parameters can potentially be optimized for shorter schedules and smaller sizes and extrapolated to longer schedules and larger sizes. 

Given the difficulty of directly optimizing the $2p$ parameters, a number of heuristic methods have been suggested which require fewer iterations. The simplest of these is the 
linear schedule, wherein the parameters at a given layer $i$ depend linearly on the fraction of the schedule $i/p$ completed thus far \cite{Shaydulin2021,Sack2021, kremenetski2021quantum, kremenetski2023quantum}. Since the initial state is a mixer eigenstate and the algorithm needs to progressively transition from the mixer to the cost Hamiltonian, the mixing angles $\beta_i$ should decrease while the cost angles $\gamma_i$ should increase throughout the schedule. This intuition leads to a simple linear interpolation:
\begin{equation}
\beta_i = a_\beta + b_\beta(1-i/p)~, \quad \gamma_i = a_\gamma + b_\gamma(i/p)~,
\end{equation}
where $a_\beta$ and $a_\gamma$ are the intercept parameters, while $b_\beta$ and $b_\gamma$ are the slope parameters. Thus only four parameters need to be optimized regardless of the number of QAOA layers $p$. While linear schedules simplify the optimization process by reducing the number of parameters, they achieve suboptimal performance in terms of approximation ratio (AR) and require high circuit depths $p$ to achieve good solutions \cite{kremenetski2021quantum,kremenetski2023quantum}. 

\begin{figure*}[ht]
    \centering
    \includegraphics[width=1\textwidth]{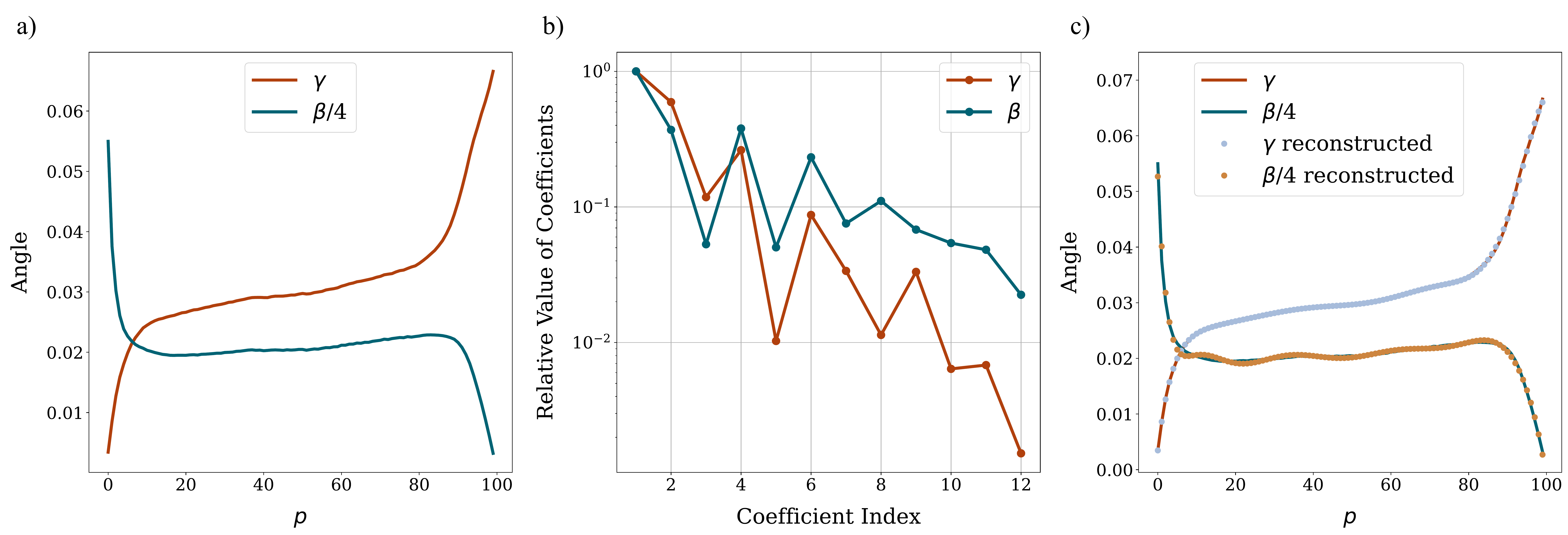}
    \caption{\label{fig:motivation}
    \textbf{Building blocks of Iterative Interpolation.} a) The parameter schedules vary smoothly with QAOA layer index. b) The coefficients of parameter schedules in the Chebyshev basis decay rapidly, showing that the first few modes have the largest contribution. c) QAOA performance with parameters reconstructed using only the first 12 coefficients is similar to that with the original schedule.}
\end{figure*}

\subsection{Fourier Interpolation for QAOA Parameters}

Based on the observation that optimized QAOA angles at depth $p+1$ are empirically close to those at depth $p$, one can extend the previously optimized parameters to higher depth. Since sine and cosine functions form an orthonormal basis on the unit interval, any schedule can be expressed as a unique combination of these functions. The choice of sine functions for $\gamma$ angles and cosine functions for $\beta$ angles is based on the boundary conditions: intuitively, the mixer angles should be maximum at the start of the schedule and decrease, while cost angles should start at zero and increase.

The ``Fourier''
strategy of Zhou et al.~\cite{zhou2020quantum} leverages this insight by performing parameter optimization in the Fourier basis. When extending from depth $p$ to $p+k$, $k$ new higher frequency components are added with zero initial amplitude, while the amplitudes of existing lower frequency components are initialized to their previously optimized values. These amplitudes are then optimized to improve the algorithm's performance. Thus, the angles are parametrized as follows:
\begin{align}
\gamma_i = \sum_{j=1}^{p} u_j \sin\left[\left(j-\frac{1}{2}\right)\left(i-\frac{1}{2} \right)\frac{\pi}{p}\right]\quad,\quad
\beta_i = \sum_{j=1}^{p} v_j \cos\left[\left(j-\frac{1}{2}\right)\left(i-\frac{1}{2} \right)\frac{\pi}{p}\right]~. 
\end{align}
The Fourier coefficients $u_j, v_j$ need to be optimized and index $i$ runs from $1$ to $p$. Note that in this method, at each step $2p$ variables still need to be optimized, but the initialization from previously optimized lower frequency components typically provides a good starting point which leads to faster convergence compared to direct parameter optimization. Nevertheless, it is still quite difficult to get beyond $p \gtrsim 100$ with this method as the optimization gets increasingly expensive.

\section{Parameter Setting by Iterative Interpolation}\label{sec:proposed_II}

We now describe the proposed Iterative Interpolation (II) parameter setting method. Our central insight is that a schedule of QAOA angles, when expressed in terms of the fraction of the schedule completed ($t = i/p$), can be viewed as a function on the unit interval $t \in [0,1]$. Any such function can be represented in the basis of orthonormal functions on the unit interval. Several families of orthogonal polynomials and functions are suitable for this decomposition, including Chebyshev polynomials, Legendre polynomials, and trigonometric functions (Fourier basis). 

Optimized QAOA schedules consistently demonstrate remarkably smooth behavior in their angle patterns. A key insight is that when such schedules are decomposed into an orthogonal function basis, the coefficients of the modes decay rapidly with only the first few modes having significant magnitude, indicating that most of the information content is captured by these low-order components. This suggests that high-fidelity reconstructions can be achieved using far fewer parameters than the original schedule length, while maintaining comparable algorithmic performance. These properties are illustrated in Fig.~\ref{fig:motivation}:
the first panel shows a smoothly varying schedule obtained for $p=100$, the second panel demonstrates the rapid decay of mode coefficients in the Chebyshev basis, and the third panel confirms that reconstruction using only the dominant modes preserves the schedule's performance. 

Consider a family of orthonormal functions on the unit interval ${f_n}$, satisfying the orthonormality condition:
\begin{equation}
\int_0^1 f_n(t)f_m(t)dt = \delta_{nm}~. 
\end{equation}
In such a basis, we can express the schedule in terms of the fraction of the schedule $i/p$, where $i$ runs from 1 to $p$ \cite{Gottlieb1977, Boyd2001}:
\begin{equation}
\gamma_i = \gamma(i/p) = \sum_{j=1}^{\mathcal{C}} u_j f_j(i/p) \quad,\quad
\beta_i = \beta(i/p) = \sum_{j=1}^{\mathcal{C}} v_j f_j(i/p)~, \label{eq:basis_representation}
\end{equation}
where $u_j$, $v_j$ are the coefficients to be optimized and $\mathcal{C}$ is the number of basis functions used in the expansion. This representation becomes exact as $\mathcal{C} \rightarrow \infty$, but in practice, a finite number of basis functions can provide an excellent approximation due to the smoothness of optimal QAOA schedules as shown in panel (c) of Fig.~\ref{fig:motivation}. By restricting ourselves to $\mathcal{C} \ll p$ coefficients, we constrain the optimization to a subspace of smooth schedules, which empirically contains high-performing solutions while dramatically reducing the optimization complexity.

A key advantage of this representation is that once the coefficients are determined, one can easily interpolate to obtain schedules for a larger value of $p$ by simply evaluating these continuous functions at the desired grid points $t_i = i/p$. This allows us to take larger steps while iteratively increasing the length of schedule. 
Importantly, the number of coefficients $\mathcal{C}$ can be much smaller than $p$, thus providing a very compact representation of the schedule.

The optimization problem now becomes one of finding optimal values for these coefficients rather than directly optimizing the angles. Efficient numerical methods exist for rapidly converting between the coefficient representation and the angles of the schedule, i.e., Eq.~\ref{eq:basis_representation}. For orthogonal polynomial bases like Chebyshev or Legendre, the coefficients can be determined by solving a linear system $Ax = b$, where $A_{ij} = f_j(i/p)$ is the evaluation of the $j$-th basis function at point $i/p$, $b$ contains the schedule angles, and $x$ contains the desired coefficients \cite{2020SciPy-NMeth, Trefethen2000}. For trigonometric functions, Fast Fourier Transform (FFT) algorithms provide an even more efficient method for computing these transformations \cite{Cooley1965}.

\begin{algorithm}[H]
    \caption{Iterative Interpolation (II)}\label{II_Alg}
    \begin{algorithmic}[1]
        \Require{$p_0$: Starting $p$ value, $\Delta p$: step increment, $p_{\text{max}}$: Max $p$ value, $\varepsilon$: Improvement threshold, $\mathcal{C}$: number of coefficients to be tuned, $\tau$: Patience , $AR'$: Desired Approximation Ratio}
        \Ensure{Optimized angles $\gamma$, $\beta$ for $p = p_{\text{max}}$}
        \State Initialize patience counter: $c_{\text{pat}} \gets 0$
        \While{$p \leq p_{\text{max}}$ \OR $AR'$ reached}
            \State Transform $\gamma^{(p)}$, $\beta^{(p)}$ to functional basis  
            \State Optimize the first $\mathcal{C}$ coefficients
            \State Compute the relative performance improvement $\delta_{\text{perf}}$
            \If{$\delta_{\text{perf}} < \varepsilon$} for $\tau$ iterations
                \State Increase the number of coefficients $\mathcal{C}$ to be tuned
            \EndIf
            \State Perform interpolation of schedule to $p + \Delta p$
        \EndWhile
    \end{algorithmic}
\end{algorithm}

The coefficients in the functional basis provide a natural framework for an iterative optimization approach as outlined in 
Algorithm \ref{II_Alg}. Starting with a low depth $p_0$, the algorithm follows these steps: First, the schedule angles are transformed into the chosen functional basis. Then, only the first $\mathcal{C}$ coefficients are optimized, reflecting our understanding that higher-order modes typically contribute less to the schedule's performance. If the relative improvement in performance (approximation ratio) falls below a threshold $\varepsilon$, this suggests we need to capture finer features of the schedule by increasing the number of coefficients being optimized. Finally, using the optimized coefficients, we can interpolate to generate a schedule at a larger depth $p + \Delta p$. This process continues iteratively until reaching the desired maximum depth $p_{\text{max}}$. To prevent premature adjustments to the number of tuned coefficients, we introduce a patience parameter $\tau$ that requires the relative performance improvement to remain below the threshold $\varepsilon$ for $\tau$ consecutive iterations before increasing $\mathcal{C}$. Note that the Fourier strategy introduced in \cite{zhou2020quantum} is a special case of this algorithm where $f_n$ are trigonometric functions. As we show in the following Section, the optimization can be made notably more efficient by choosing $\mathcal{C} \ll p$ coefficients to optimize and setting $\Delta p > 1$. 

In our experiments, we use the BOBYQA algorithm for optimizing the  coefficients as it is well-suited for derivative-free optimization and has demonstrated robust performance in this context. However, our approach is not restricted to BOBYQA; other optimizers such as COBYLA or Nelder-Mead can also be employed effectively, and we expect similar results with these alternatives \cite{singer2009nelder,conn2009introduction,powell2009bobyqa}.

\subsection{Theoretical Motivation}

A central theoretical motivation for our approach is the classical result that smooth or Lipschitz continuous functions can be efficiently approximated by low-degree polynomials or other orthogonal basis expansions. The QAOA parameter schedules when viewed as functions on the unit interval are observed to be smooth in practice~\cite{shaydulin2023evidence}. 

A quantitative bound on the error of best uniform polynomial approximation for Lipschitz continuous functions, can be obtained by Jackson's Theorem~\cite{Jackson2005-ir, Cheney1998-pn}

\begin{theorem}[Jackson's Theorem for Lipschitz Functions]
Let $f: [0,1] \to \mathbb{R}$ be Lipschitz continuous with constant $L$, i.e.,
$$
|f(x) - f(y)| \leq L |x - y| \quad \forall x, y \in [0,1].
$$
Then, $\forall n \geq 1$, there exists a polynomial $P_n$ of degree at most $n$ such that
$$
\max_{x \in [0,1]} |f(x) - P_n(x)| \leq C \frac{L}{n}
$$
where $C$ is an absolute constant.
\end{theorem}
Note that when the function is $f \in C^2[0,1]$, the rate of approximation is $\mathcal{O}(1/n^2)$, and thus decays faster than $\mathcal{O}(1/n)$ for $f \in C^1[0,1]$. This result implies that the number of basis coefficients required to approximate a Lipschitz continuous schedule to within a uniform error $\varepsilon$ is at most $n \geq C L / \varepsilon$. In other words, the smoother the schedule (i.e., the smaller the Lipschitz constant $L$), the fewer coefficients are needed for accurate representation. This provides a theoretical explanation for the empirical observation that QAOA schedules can be well-approximated with a small number of basis functions as shown in Figure~\ref{fig:motivation}, and justifies our use of truncated orthogonal expansions (e.g., Chebyshev or Fourier bases) in the optimization process.

In turn, the expectation value of the cost function is Lipschitz continuous in terms of the QAOA angles. This can be derived based on the following theorem~\cite{Sweke2020}

\begin{theorem}[Lipschitz Continuity of Expectation Values]\label{thm:lips_cont1}
Consider a function $f:[0,2\pi]^M\mapsto\mathbb{R}$ defined by $f(\vec{\theta})=\langle \psi|U^\dagger(\vec{\theta})~O~U(\vec{\theta})|\psi\rangle$, where $|\psi\rangle$ is an arbitrary state in $\mathbb{C}^D$ for some finite $D$, $U(\vec{\theta})$ is a quantum circuit consisting of an arbitrary but finite number of fixed gates, and $M$ parameterized gates $U_j(\theta_j) = \mathrm{exp}(-i(\theta_j + c_j)H_j)$, with $H_j$ Hermitian. For any observable $O$ and any set of Hermitian operators $\{H_j\, |\, j\in[M]\}$, the function $f(\vec{\theta})$ is $L$-Lipschitz with $L \leq 2\sqrt{M}\,\max_j \Vert O \Vert_2~ \Vert H_j \Vert_2$.
\end{theorem}

By picking $\vec{\theta} = (\bm\gamma, \bm\beta)$ with the corresponding Hermitian operators $H_C, H_B$, we obtain the following corollary  

\begin{corollary}[Lipschitz Continuity of QAOA Expectation]\label{corollary:lips_qaoa}

The normalized expectation value of $H_C$ in the QAOA state $\vert \bm\gamma, \bm\beta \rangle$ after $p$-layers given by $f(\bm\gamma, \bm\beta) = \langle \bm\gamma, \bm\beta \vert H_C \vert \bm\gamma, \bm\beta \rangle / \Vert H_C \Vert_2 $ is 
$L$-Lipschitz with $L \leq 2\sqrt{2 p}\,\max(\Vert H_B \Vert_2~,~ \Vert H_C \Vert_2)$.
\end{corollary}

We can now combine the Lipschitz continuity of the QAOA expectation value with Jackson's theorem to derive a bound on the number of coefficients needed to achieve a certain relative approximation error in the cost function expectation value. 

\begin{corollary}[Coefficient Complexity for Relative Accuracy]\label{corollary:coeff_complexity}
Let $p\in\mathbb{N}$ be the QAOA depth, and suppose the QAOA angles are generated by schedules $g,b:[0,1]\to\mathbb{R}$ via $\gamma_k=g(k/p)$ and $\beta_k=b(k/p)$ for $k=1,\dots,p$. Assume $g$ and $b$ are Lipschitz with constants $L_g$ and $L_b$, and set $L_* := \sup\{L_g, L_b\} = \max(L_g,L_b)$. Consider truncated orthogonal basis expansions (e.g., Chebyshev or Fourier) of degree $n$ for both $g$ and $b$, producing approximations $g_n,b_n$ with uniform errors
$$
\|g-g_n\|_\infty \le \frac{C\,L_g}{n},\qquad \|b-b_n\|_\infty \le \frac{C\,L_b}{n}
$$
for an absolute constant $C>0$ as in Jackson's Theorem. Define the normalized cost expectation
$$
f(\bm\gamma,\bm\beta)=\frac{\langle \bm\gamma,\bm\beta \vert H_C \vert \bm\gamma,\bm\beta\rangle}{\|H_C\|_2}.
$$
Then, we can guarantee a relative accuracy $\varepsilon>0$ in the normalized expectation 
$$
\big|f(\bm\gamma,\bm\beta)-f(\bm\gamma^{(n)},\bm\beta^{(n)})\big| \le \varepsilon
$$ with a shared degree of at most 
$$
n \;\le\; \frac{2\sqrt{2}\,C\,p\,L_* \,\max(\|H_B\|_2,\|H_C\|_2)}{ \varepsilon}.
$$
\end{corollary}

As expected, schedules which are longer (higher $p$) or have larger Lipschitz constant ($L_*$), require greater number of coefficients to achieve the same relative error. Furthermore, as the amount of change in the QAOA state at a given step is controlled by the cost and mixer Hamiltonians, the number of coefficients grows with their spectral norm.

Regarding the convergence of our iterative optimization procedure, we note that the underlying optimization landscape is generally non-convex. However, standard results from nonlinear optimization, guarantee local convergence under mild regularity conditions. Specifically, if the objective function (e.g., the QAOA cost as a function of the basis coefficients) is continuously differentiable, then local derivative-free optimizers such as BOBYQA or COBYLA will converge to a stationary point. This is formalized by the following statement:

\begin{theorem}[Convergence of Trust-Region Derivative-Free Methods~\cite{Conn2009}]
Let $F: D \to \mathbb{R}$ be continuously differentiable on a compact set $D \subset \mathbb{R}^n$. Suppose a trust-region derivative-free algorithm (such as BOBYQA) is applied, and the sequence of iterates $\{x_k\}$ is generated such that the trust-region radius $\Delta_k \to 0$. Then, every limit point of $\{x_k\}$ is a stationary point of $F$.
\end{theorem}

In summary, the combination of a bound on the number of coefficients for a fixed relative error with local convergence guarantees for derivative-free optimization provides a theoretical foundation for our method. The number of basis coefficients required for a given accuracy is controlled by the smoothness of the schedule, and the optimization procedure is guaranteed to converge locally under standard assumptions. This theoretical underpinning complements our empirical results presented in the following section.




\begin{figure*}[!htb]
    \centering
    \includegraphics[width=1\textwidth]{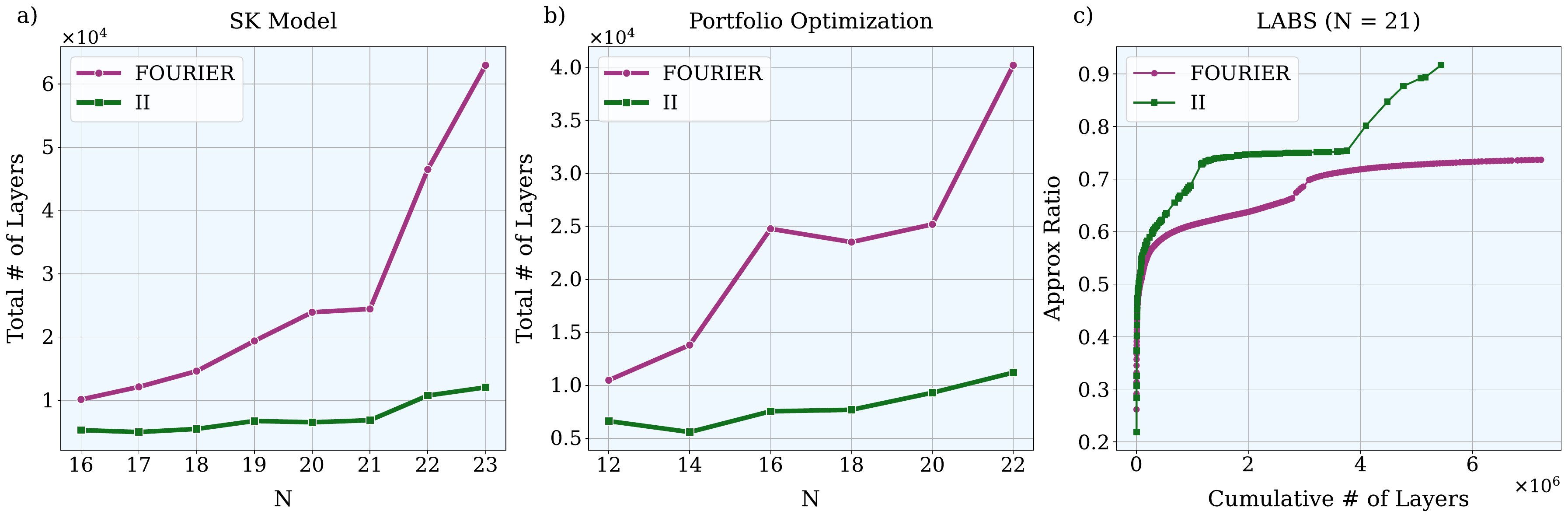}
    \caption{\label{prob_eval}
    \textbf{Performance of Chebhyshev II in comparison to the Fourier method}. Across different problem sizes considered, II performs significantly better when compared to the Fourier method. a) SK model: II achieves $50\%$ overlap with the ground state using $3.3\times$ fewer total layers, estimated as an average across of $N$ of the ratio of median total number of layers for Fourier to that of II. b) Portfolio optimization: II consistently reaches approximation ratios $\geq 0.9$ with $2.8\times$ fewer total layers than Fourier. c) LABS: II achieves better approximation ratios ($\simeq 0.95$) while Fourier plateaus below $0.74$, despite using substantially fewer total layers. The values annotated in Figs. a) and b) correspond to the median values across different seeds for each problem instance.}
\end{figure*}

\section{Results}\label{sec:results}

\begin{figure*}[htb]
    \centering
    \includegraphics[width=1\textwidth]{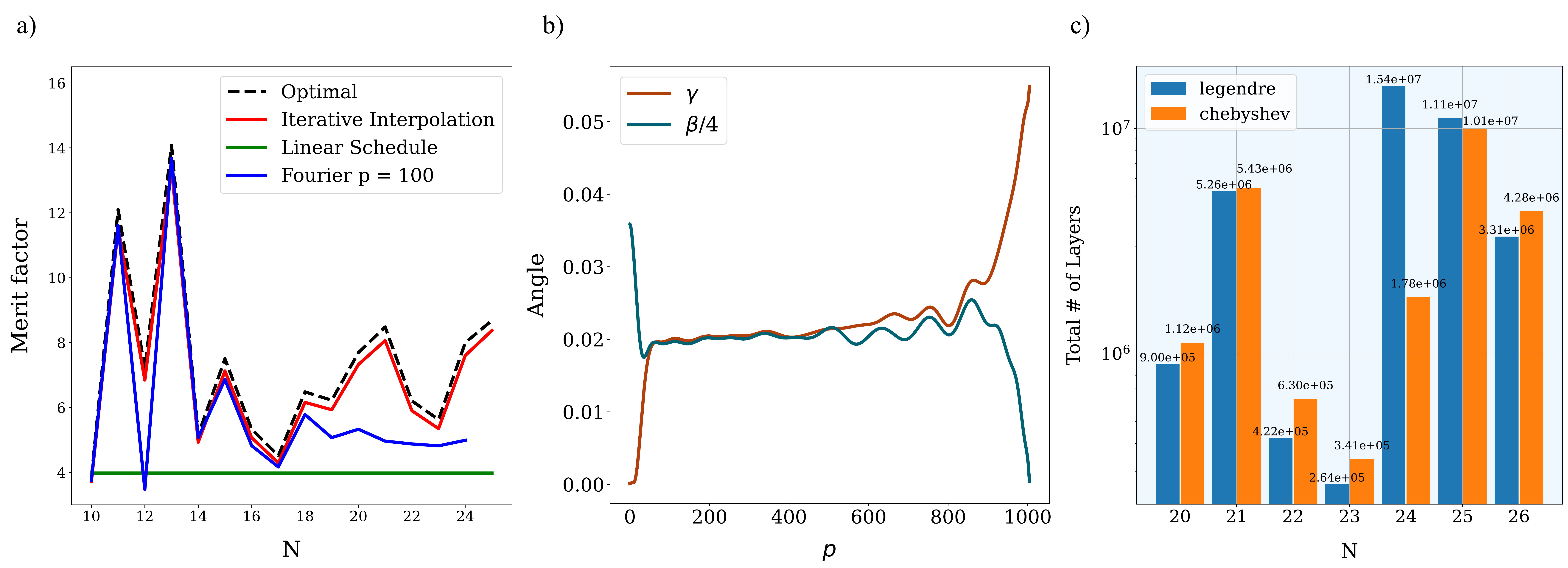}
    \caption{\label{labs_results}
    \textbf{II obtains the optimal values for LABS}. a) The comparison across different QAOA parameter schedules for solving the LABS problem. The II schedule attains the optimal MFs for the $N$ values considered, while the Fourier and Linear schedules fail to do so. b) The II schedule corresponding to $N=25$, a $p=1005$ and AR$=0.965$. c) The impact of the basis function choice on performance. Although the choice of basis can be optimized for each instance, the overall performance across different $N$ remains comparable for both choices. 
    }
\end{figure*}

We analyze the performance of iterative interpolation (II) in comparison to the established technique of the Fourier method with $\Delta p=1$. We define {Total $\#$ of Layers} to be $\sum_i i {f^i_{eval}}$, where ${f^i_{eval}}$ is the number of functional evaluations at the $i^{\text{th}}$ QAOA layer. The functional evaluation implies the number of times the optimization routine evaluates the QAOA objective for a given problem. Hereon, we will refer to {Total $\#$ of Layers} given by $\sum_i i {f^i_{eval}}$ as \tnl. The II method consistently exhibits better performance evidenced by the lower median values in the \tnl. It is also noteworthy that in the case of the LABS problem, the approximation ratio obtained by executing the Fourier method is significantly lower even when the \tnl executed is much higher. The numerical results are obtained by state-vector simulations using the QOKIT~\cite{Lykov2023} library running on a NVIDIA A10 GPU with 24 GB of VRAM.


Figure \ref{prob_eval} provides a visual comparison of II in the Chebyshev versus the Fourier method across our three benchmark problems. The results for the following case studies are summarized below:

\begin{itemize}
    \item \textbf{SK model}: The panel \textit{a} from Fig.~\ref{prob_eval} presents the superior performance of II when compared to the Fourier technique for the SK model. The results pertain to executing II and {Fourier} for $50$ seeds corresponding to each $N$. The $p_{max}$ and $\Delta p$ were set to be $2000$ and $5$ respectively. The result was said to be approximately optimal when the optimized angles corresponding to a method obtained an overlap of $50\%$ with the exact ground state. The II method obtained a $3.3\times$ median improvement over the Fourier method, when averaged across all $N$ considered. 
    \item \textbf{Portfolio optimization}: The panel \textit{b} from Fig.~\ref{prob_eval} is the result corresponding to solving PO problem using II and comparing its performance against that of the Fourier technique. $10$ seeds corresponding to each $N$ were used as the problem to test the performance of the two methods. The approximation ratio $\geq 0.9$ was chosen to be the optimality condition, where an approximation ratio $=1$ corresponds to solving the problem exactly. The II method obtained a $2.8\times$ median improvement over the Fourier method, when averaged across all $N$ considered.  
    \item \textbf{LABS}: The panel \textit{c} in Fig.~\ref{prob_eval} describes the improvement in AR as a function of the cumulative number of layers for the LABS problem of size $N=21$. For this specific case, we consider the Legendre basis, patience $\tau = 5$, and step size $\Delta p = 5$. Executing {II} yields a better AR at the cost of using significantly lower number of layers in comparison to the Fourier method. It is also worth noting that the Fourier method fails to go beyond an $\text{AR}>0.74$, while II achieves an $\text{AR}>0.95$. 
    
    Figure \ref{labs_results} shows the performance of II for the LABS problem across different $N$ values. As demonstrated in panel \textit{a}, II achieves optimal merit factors across all tested problem sizes, outperforming both Fourier and Linear schedules. While Linear schedules use only four parameters regardless of depth which limits their expressivity, Fourier schedules become computationally prohibitive at large depths as they optimize all $2p$ parameters. In contrast, II reaches significantly larger depths ($p>1000$) by optimizing only a small set of basis coefficients, enabling it to find higher-quality solutions than the other approaches. 
    
    Panel \textit{b} of Figure \ref{labs_results} displays the parameter schedule obtained using II for $N=25$ with $p=1005$, achieving AR$=0.965$. Notably, although we didn't explicitly enforce boundary conditions, the characteristic pattern of $\gamma$ increasing from zero and $\beta$ decreasing to zero naturally emerges during optimization. Panel \textit{c} illustrates that while the specific choice of basis functions can be tuned, the performance of different basis types is comparable - suggesting that II's effectiveness stems primarily from the adaptive coefficient selection rather than the particular basis employed. 
\end{itemize}


\subsection{Growth of QAOA depth required to solve Sherrington-Kirkpatrick and LABS problems}
Previous studies have focused on analyzing the scaling of the time to solution of QAOA with constant depth~\cite{shaydulin2023evidence,boulebnane2022solving}. Concretely, if QAOA with a fixed depth $p$ produces overlap $\delta^*$ with the ground state of the problem Hamiltonian, then QAOA time to solution is $O\left(\frac{1}{\delta^*}\right)$ and $O\left(\frac{1}{\sqrt{\delta^*}}\right)$ with amplitude amplification. An alternative and less studied setting is to allow QAOA depth to grow with problem size and analyze the rate of growth to achieve constant $\delta^*$. It is possible that the scaling in the latter setting would be more favorable, motivating its study. However, evaluating optimized QAOA performance in this regime requires optimizing a large number of parameters, which was prohibitively expensive with prior techniques. 

\begin{figure*}[htb]
    \begin{subfigure}[b]{0.45\columnwidth}
        \subcaption{}
        \includegraphics[width=\columnwidth]{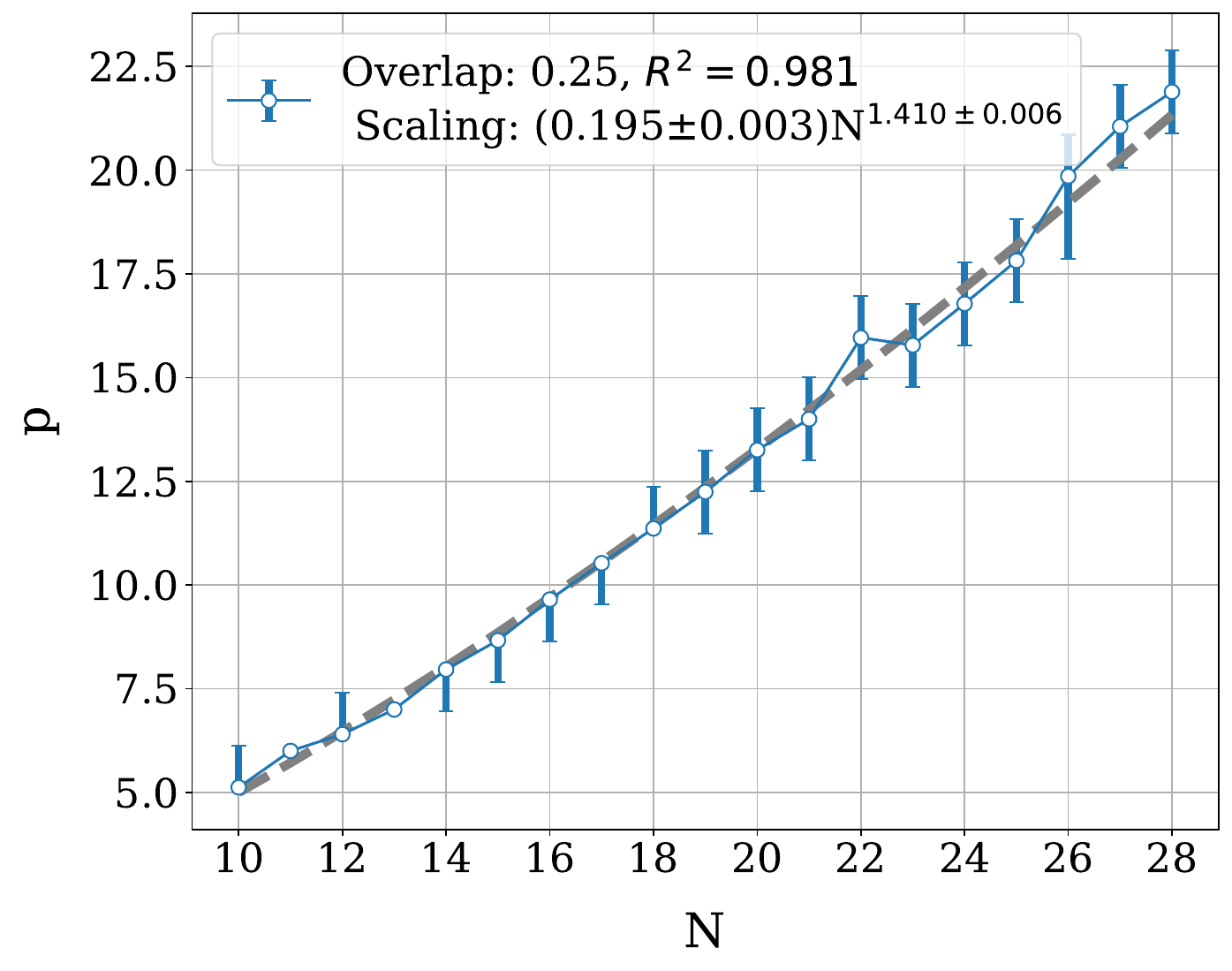}
        \label{fig:prob_scaling_5a}
    \end{subfigure}
    \begin{subfigure}[b]{0.45\columnwidth}
        \subcaption{}
        \includegraphics[width=\columnwidth]{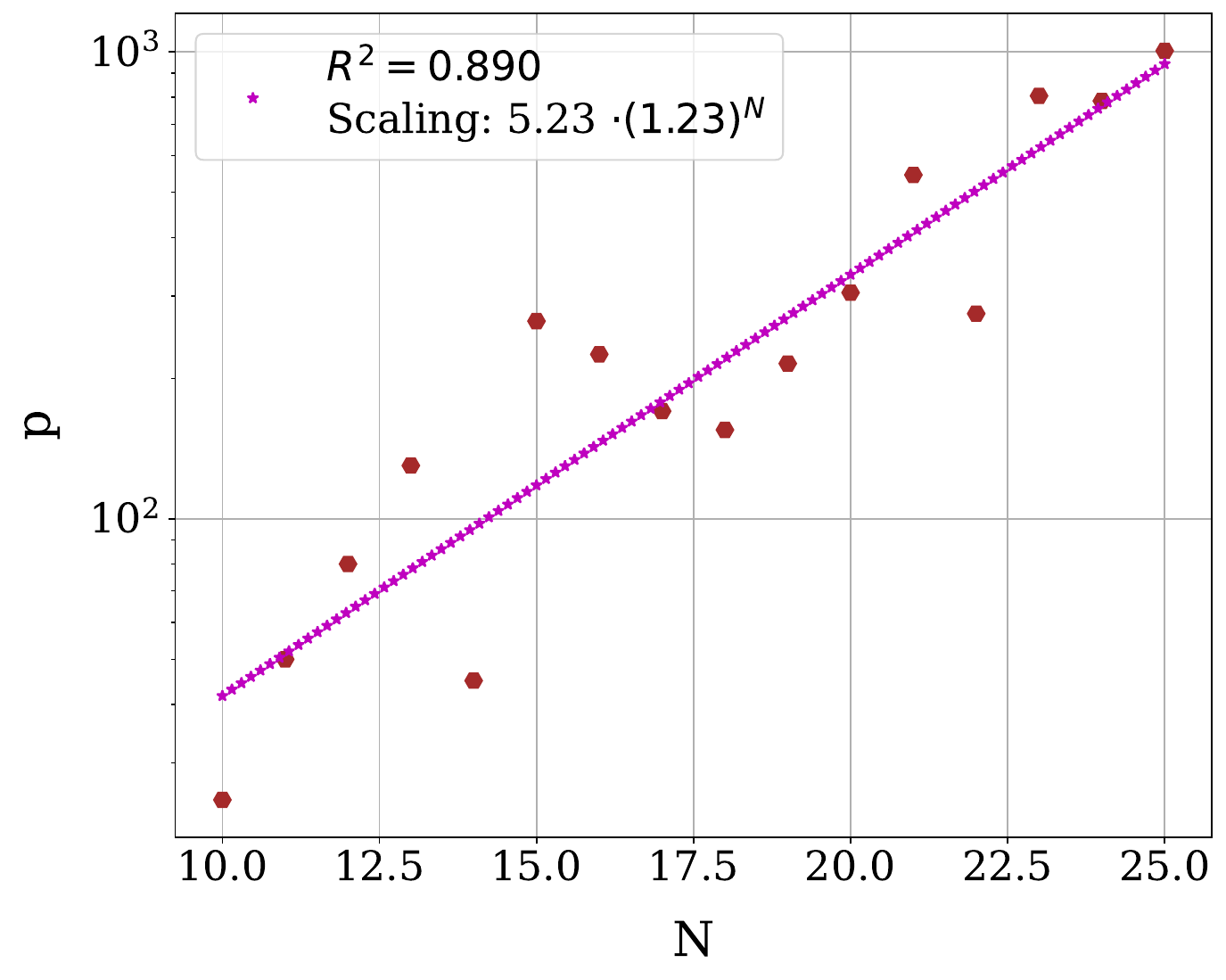}
        \label{fig:prob_scaling_5b}
        
    \end{subfigure}
    \caption{\label{prob_scaling}
    The depth $p$ required to achieve a fixed overlap with the optimal state for the SK model (a) and LABS (b). For the SK model, we fit the depth to a polynomial function in $N$ and report a high goodness-of-fit. However, we are not able to make rigorous claims about scaling. For LABS, the depth appears to grow exponentially as a function of $N$, while for the SK model the growth of $p$ appears slower.}
\end{figure*}

Fig.~\ref{prob_scaling} shows the scaling of QAOA depth with system size for the SK model and LABS with parameters optimized using II. For the SK model, we randomly generate 600 instances of the problem for system sizes in the range $[10,28]$ and run II with a maximum depth of $p=150$. We report the depth required to reach a $25\%$ overlap with the optimal state and present the median depth required, as well as a $5\%$ confidence interval about the median. For LABS problem, we report the depth required to achieve $0.9$ approximation ratio.

For the SK model, we fit the depth to a polynomial function in system size in and find the relation $p=0.19\cdot  N^{1.41}$, with a goodness-of-fit ($\text{R}^2$) value of $.98$ (visualized in Fig.~\ref{prob_scaling}a). We observe better goodness-of-fit values for polynomial fit as compared to the exponential fit. However, we do not make a claim of polynomial scaling of QAOA time to solution for SK due to the limited scale of our numerical experiments. We remark that, the message-passing algorithm of \cite{montanari2019} is able to obtain a fixed constant \emph{relative error} of $\epsilon$ on average in time $\mathcal{O}(C(\epsilon)N^2)$, analogous to the results for QAOA in \cite{basso2021quantum, Basso2022} (as these results have constant number of QAOA layers, where each layer involves $\mathcal{O}(n^2)$ gates, assuming all-to-all connectivity). It is unclear, however, what the required runtime to obtain a fixed \emph{absolute error}. We may view our results as a step towards exploring QAOA in this regime. We present additional results on the scaling for different choices of target overlap in Appendix~\ref{sec:appendix_extra_SK_numerics}. 

Compared to the SK model, the QAOA depth for LABS grows much faster (see Fig.~\ref{prob_scaling}b), though the noisiness of the data makes it hard to make any strong claim about the scaling. We remark, however, that an exponential fit gives the scaling of $1.23^N$, which is close to the $(1.21\pm 0.2)^N$ scaling reported in Ref.~\cite{shaydulin2023evidence} of QAOA with $p=12$ combined with quantum minimum finding. 

\section{Discussion}\label{sec:discussion}

This work demonstrates that expressing QAOA parameter schedules as a small number of smooth basis functions provides a powerful framework for optimization. Our approach enables efficient representation of schedules with hundreds of layers, enabling scaling QAOA to regimes inaccessible by existing methods. By adaptively selecting and optimizing only the most significant basis coefficients, our iterative interpolation technique dramatically reduces optimization complexity while improving solution quality.

Note that we did not undertake simulations of our method with noisy quantum circuits and it will be important to understand how the method fares compared to the traditional methods in the presence of noise. Furthermore, experiments on hardware will demonstrate whether the advantages observed in simulation persist in real-world quantum devices. Ultimately, validating the efficacy of our method under realistic conditions will be essential for its adoption in practical quantum optimization tasks.

Our findings open several promising research directions. The technique could be extended to optimize continuous-time quantum annealing schedules, such as those used in neutral atom platforms for maximum independent set problems~\cite{ebadi:22,andrist:23}. Additionally, we observe a stark contrast in the scaling behavior between different problem classes: the depth required to solve Sherrington-Kirkpatrick instances in the average case appears to grow polynomially with problem size, while LABS exhibits exponential scaling. This pronounced difference in computational requirements invites deeper theoretical analysis into which structural properties determine QAOA's efficacy and why certain problem classes are fundamentally more amenable to quantum optimization approaches than others \cite{boulebnane2025}.

The dramatic performance improvement achieved by our method highlights the importance of efficient parameter optimization strategies for QAOA.
As quantum hardware continues to advance, techniques like iterative interpolation that exploit structure in optimal schedules will be crucial for realizing quantum advantage in practical optimization tasks.




\section*{Data Availability}

The full data presented in this work is available at  \href{https://doi.org/10.5281/zenodo.15120519}{https://doi.org/10.5281/zenodo.15120519}.

\vspace{3mm}
\section*{Code Availability}

The code implementing the proposed parameter setting method is available at \href{https://github.com/jpmorganchase/QOKit/blob/main/examples/QAOA_iterative_interpolation_SK.ipynb}{QOKIT}~\cite{Lykov2023}.

\bibliographystyle{unsrt}
\bibliography{references}

@article{farhi2014quantum,
  title={A quantum approximate optimization algorithm},
  author={Farhi, Edward and Goldstone, Jeffrey and Gutmann, Sam},
  journal={arXiv preprint arXiv:1411.4028},
  year={2014}
}

@article{zhou2020quantum,
  author={Zhou, Leo and Wang, Sheng-Tao and Choi, Soonwon and Pichler, Hannes and Lukin, Mikhail D},
  title={Quantum approximate optimization algorithm: Performance, mechanism, and implementation on near-term devices},
  journal={Physical Review X},
  volume={10},
  number={2},
  pages={021067},
  year={2020},
  publisher={APS}
}

@article{shaydulin2023evidence,
  title={Evidence of scaling advantage for the quantum approximate optimization algorithm on a classically intractable problem},
  author={Shaydulin, Ruslan and Li, Changhao and Chakrabarti, Shouvanik and DeCross, Matthew and Herman, Dylan and Kumar, Niraj and Larson, Jeffrey and Lykov, Danylo and Minssen, Pierre and Sun, Yue and others},
  journal={arXiv preprint arXiv:2308.02342},
  year={2023}
}

@article{basso2021quantum,
  title={The quantum approximate optimization algorithm at high depth for MaxCut on large-girth regular graphs and the Sherrington-Kirkpatrick model},
  author={Basso, Joao and Farhi, Edward and Marwaha, Kunal and Villalonga, Benjamin and Zhou, Leo},
  journal={arXiv preprint arXiv:2110.14206},
  year={2021}
}

@article{farhi2022quantum,
  title={The quantum approximate optimization algorithm and the sherrington-kirkpatrick model at infinite size},
  author={Farhi, Edward and Goldstone, Jeffrey and Gutmann, Sam and Zhou, Leo},
  journal={Quantum},
  volume={6},
  pages={759},
  year={2022},
  publisher={Verein zur F{\"o}rderung des Open Access Publizierens in den Quantenwissenschaften}
}

@article{boulebnane2022solving,
  title={Solving boolean satisfiability problems with the quantum approximate optimization algorithm},
  author={Boulebnane, Sami and Montanaro, Ashley},
  journal={arXiv preprint arXiv:2208.06909},
  year={2022}
}

@misc{montanari2019,
      title={Optimization of the Sherrington-Kirkpatrick Hamiltonian}, 
      author={Andrea Montanari},
      year={2019},
      eprint={1812.10897},
      archivePrefix={arXiv},
      primaryClass={math.PR},
      url={https://arxiv.org/abs/1812.10897}, 
}

@inproceedings{galda2021transferability,
  title={Transferability of optimal QAOA parameters between random graphs},
  author={Galda, Alexey and Liu, Xiaoyuan and Lykov, Danylo and Alexeev, Yuri and Safro, Ilya},
  booktitle={2021 IEEE International Conference on Quantum Computing and Engineering (QCE)},
  pages={171--180},
  year={2021},
  organization={IEEE}
}

@article{shaydulin2023parameter,
  title={Parameter transfer for quantum approximate optimization of weighted maxcut},
  author={Shaydulin, Ruslan and Lotshaw, Phillip C and Larson, Jeffrey and Ostrowski, James and Humble, Travis S},
  journal={ACM Transactions on Quantum Computing},
  volume={4},
  number={3},
  pages={1--15},
  year={2023},
  publisher={ACM New York, NY}
}

@article{sureshbabu2024parameter,
  title={Parameter setting in quantum approximate optimization of weighted problems},
  author={Sureshbabu, Shree Hari and Herman, Dylan and Shaydulin, Ruslan and Basso, Joao and Chakrabarti, Shouvanik and Sun, Yue and Pistoia, Marco},
  journal={Quantum},
  volume={8},
  pages={1231},
  year={2024},
  publisher={Verein zur F{\"o}rderung des Open Access Publizierens in den Quantenwissenschaften}
}

@article{brandao2018fixed,
  title={For fixed control parameters the quantum approximate optimization algorithm's objective function value concentrates for typical instances},
  author={Brandao, Fernando GSL and Broughton, Michael and Farhi, Edward and Gutmann, Sam and Neven, Hartmut},
  journal={arXiv preprint arXiv:1812.04170},
  year={2018}
}

@article{Pagano_2020,
   title={Quantum approximate optimization of the long-range Ising model with a trapped-ion quantum simulator},
   volume={117},
   ISSN={1091-6490},
   url={http://dx.doi.org/10.1073/pnas.2006373117},
   DOI={10.1073/pnas.2006373117},
   number={41},
   journal={Proceedings of the National Academy of Sciences},
   publisher={Proceedings of the National Academy of Sciences},
   author={Pagano, Guido and Bapat, Aniruddha and Becker, Patrick and Collins, Katherine S. and De, Arinjoy and Hess, Paul W. and Kaplan, Harvey B. and Kyprianidis, Antonis and Tan, Wen Lin and Baldwin, Christopher and Brady, Lucas T. and Deshpande, Abhinav and Liu, Fangli and Jordan, Stephen and Gorshkov, Alexey V. and Monroe, Christopher},
   year={2020},
   month=oct, pages={25396–25401} }

@article{brady2021behavior,
  title={Behavior of analog quantum algorithms},
  author={Brady, Lucas T and Kocia, Lucas and Bienias, Przemyslaw and Bapat, Aniruddha and Kharkov, Yaroslav and Gorshkov, Alexey V},
  journal={arXiv preprint arXiv:2107.01218},
  year={2021}
}

@article{Hogg2000,
  title = {Quantum optimization},
  volume = {128},
  ISSN = {0020-0255},
  DOI = {10.1016/s0020-0255(00)00052-9},
  number = {3–4},
  journal = {Information Sciences},
  publisher = {Elsevier BV},
  author = {Hogg,  Tad and Portnov,  Dmitriy},
  year = {2000},
  month = oct,
  pages = {181–197}
}

@article{Hogg2000search,
  title = {Quantum search heuristics},
  volume = {61},
  ISSN = {1094-1622},
  url = {http://dx.doi.org/10.1103/PhysRevA.61.052311},
  DOI = {10.1103/physreva.61.052311},
  number = {5},
  journal = {Physical Review A},
  publisher = {American Physical Society (APS)},
  author = {Hogg,  Tad},
  year = {2000},
  month = apr 
}

@inproceedings{Shaydulin2023npgeq,
  title = {QAOA with $N\cdot p\geq 200$},
  url = {http://dx.doi.org/10.1109/QCE57702.2023.00121},
  DOI = {10.1109/qce57702.2023.00121},
  booktitle = {2023 IEEE International Conference on Quantum Computing and Engineering (QCE)},
  publisher = {IEEE},
  author = {Shaydulin,  Ruslan and Pistoia,  Marco},
  year = {2023},
  month = sep,
  pages = {1074–1077}
}

@incollection{Pelofske2023,
  doi = {10.1007/978-3-031-32041-5_13},
  outurl = {https://doi.org/10.1007/978-3-031-32041-5_13},
  year = {2023},
  publisher = {Springer Nature Switzerland},
  pages = {240--258},
  author = {Elijah Pelofske and Andreas B\"{a}rtschi and Stephan Eidenbenz},
  title = {Quantum Annealing vs. {QAOA}: 127 Qubit Higher-Order Ising Problems on~{NISQ} Computers},
  booktitle = {Lecture Notes in Computer Science}
}

@article{Pelofske2024,
  title = {Scaling whole-chip QAOA for higher-order ising spin glass models on heavy-hex graphs},
  volume = {10},
  ISSN = {2056-6387},
  url = {http://dx.doi.org/10.1038/s41534-024-00906-w},
  DOI = {10.1038/s41534-024-00906-w},
  number = {1},
  journal = {npj Quantum Information},
  publisher = {Springer Science and Business Media LLC},
  author = {Pelofske,  Elijah and B\"{a}rtschi,  Andreas and Cincio,  Lukasz and Golden,  John and Eidenbenz,  Stephan},
  year = {2024},
  month = nov 
}

@article{2409.12104,
Author = {Zichang He and David Amaro and Ruslan Shaydulin and Marco Pistoia},
Title = {Performance of Quantum Approximate Optimization with Quantum Error Detection},
Year = {2024},
journal = {arXiv:2409.12104},
}

@article{ICCAD_qaoapara,
  title={Parameter Setting Heuristics Make the Quantum Approximate Optimization Algorithm Suitable for the Early Fault-Tolerant Era},
  author={He, Zichang and Shaydulin, Ruslan and Herman, Dylan and Li, Changhao and Sureshbabu, Shree Hari and Pistoia, Marco},
  journal={arXiv preprint arXiv:2408.09538},
  year={2024},
  url={https://arxiv.org/abs/2408.09538}, 
}

@article{2411.17442,
Author = {Sami Boulebnane and Maria Ciudad-Alañón and Lana Mineh and Ashley Montanaro and Niam Vaishnav},
Title = {Applying the quantum approximate optimization algorithm to general constraint satisfaction problems},
Year = {2024},
journal = {arXiv:2411.17442},
}

@article{2411.04979,
Author = {Ashley Montanaro and Leo Zhou},
Title = {Quantum speedups in solving near-symmetric optimization problems by low-depth QAOA},
Year = {2024},
journal = {arXiv:2411.04979},
}

@article{Sack2021,
  title = {Quantum annealing initialization of the quantum approximate optimization algorithm},
  volume = {5},
  ISSN = {2521-327X},
  url = {http://dx.doi.org/10.22331/q-2021-07-01-491},
  DOI = {10.22331/q-2021-07-01-491},
  journal = {Quantum},
  publisher = {Verein zur Forderung des Open Access Publizierens in den Quantenwissenschaften},
  author = {Sack,  Stefan H. and Serbyn,  Maksym},
  year = {2021},
  month = jul,
  pages = {491}
}

@article{Sud2024,
  title = {Parameter-setting heuristic for the quantum alternating operator ansatz},
  volume = {6},
  ISSN = {2643-1564},
  url = {http://dx.doi.org/10.1103/PhysRevResearch.6.023171},
  DOI = {10.1103/physrevresearch.6.023171},
  number = {2},
  journal = {Physical Review Research},
  publisher = {American Physical Society (APS)},
  author = {Sud,  James and Hadfield,  Stuart and Rieffel,  Eleanor and Tubman,  Norm and Hogg,  Tad},
  year = {2024},
  month = may 
}

@article{Golay1972,

  author={Golay, M.},

  journal={IEEE Transactions on Information Theory}, 

  title={A class of finite binary sequences with alternate auto-correlation values equal to zero (Corresp.)}, 

  year={1972},

  volume={18},

  number={3},

  pages={449-450},

  doi={10.1109/TIT.1972.1054797}}

@article{Pasha2000,
  title = {Bi-alphabetic pulse compression radar signal design},
  volume = {25},
  ISSN = {0973-7677},
  url = {http://dx.doi.org/10.1007/BF02703629},
  DOI = {10.1007/bf02703629},
  number = {5},
  journal = {Sadhana},
  publisher = {Springer Science and Business Media LLC},
  author = {Pasha,  I. A. and Moharir,  P. S. and Rao,  N. Sudarshan},
  year = {2000},
  month = oct,
  pages = {481–488}
}

@article{Packebusch2016,
  title = {Low autocorrelation binary sequences},
  volume = {49},
  ISSN = {1751-8121},
  url = {http://dx.doi.org/10.1088/1751-8113/49/16/165001},
  DOI = {10.1088/1751-8113/49/16/165001},
  number = {16},
  journal = {Journal of Physics A: Mathematical and Theoretical},
  publisher = {IOP Publishing},
  author = {Packebusch,  Tom and Mertens,  Stephan},
  year = {2016},
  month = mar,
  pages = {165001}
}

@article{Markowitz1952,
  title = {PORTFOLIO SELECTION},
  volume = {7},
  ISSN = {1540-6261},
  url = {http://dx.doi.org/10.1111/j.1540-6261.1952.tb01525.x},
  DOI = {10.1111/j.1540-6261.1952.tb01525.x},
  number = {1},
  journal = {The Journal of Finance},
  publisher = {Wiley},
  author = {Markowitz,  Harry},
  year = {1952},
  month = mar,
  pages = {77–91}
}

@article{DeMiguel2007,
  title = {Optimal Versus Naive Diversification: How Inefficient is the 1/N Portfolio Strategy?},
  volume = {22},
  ISSN = {1465-7368},
  url = {http://dx.doi.org/10.1093/rfs/hhm075},
  DOI = {10.1093/rfs/hhm075},
  number = {5},
  journal = {Review of Financial Studies},
  publisher = {Oxford University Press (OUP)},
  author = {DeMiguel,  Victor and Garlappi,  Lorenzo and Uppal,  Raman},
  year = {2007},
  month = dec,
  pages = {1915–1953}
}

@article{Black1992,
  title = {Global Portfolio Optimization},
  volume = {48},
  ISSN = {1938-3312},
  url = {http://dx.doi.org/10.2469/faj.v48.n5.28},
  DOI = {10.2469/faj.v48.n5.28},
  number = {5},
  journal = {Financial Analysts Journal},
  publisher = {Informa UK Limited},
  author = {Black,  Fischer and Litterman,  Robert},
  year = {1992},
  month = sep,
  pages = {28–43}
}

@article{Brandhofer2022,
  title = {Benchmarking the performance of portfolio optimization with QAOA},
  volume = {22},
  ISSN = {1573-1332},
  url = {http://dx.doi.org/10.1007/s11128-022-03766-5},
  DOI = {10.1007/s11128-022-03766-5},
  number = {1},
  journal = {Quantum Information Processing},
  publisher = {Springer Science and Business Media LLC},
  author = {Brandhofer,  Sebastian and Braun,  Daniel and Dehn,  Vanessa and Hellstern,  Gerhard and H\"{u}ls,  Matthias and Ji,  Yanjun and Polian,  Ilia and Bhatia,  Amandeep Singh and Wellens,  Thomas},
  year = {2022},
  month = dec 
}

@article{Buonaiuto2023,
  title = {Best practices for portfolio optimization by quantum computing,  experimented on real quantum devices},
  volume = {13},
  ISSN = {2045-2322},
  url = {http://dx.doi.org/10.1038/s41598-023-45392-w},
  DOI = {10.1038/s41598-023-45392-w},
  number = {1},
  journal = {Scientific Reports},
  publisher = {Springer Science and Business Media LLC},
  author = {Buonaiuto,  Giuseppe and Gargiulo,  Francesco and De Pietro,  Giuseppe and Esposito,  Massimo and Pota,  Marco},
  year = {2023},
  month = nov 
}

@article{Yuan_2026,
doi = {10.1088/2058-9565/ae4a48},
url = {https://doi.org/10.1088/2058-9565/ae4a48},
year = {2026},
month = {mar},
publisher = {IOP Publishing},
volume = {11},
number = {2},
pages = {025034},
author = {Yuan, Haomu and Long, Christopher K and Lepage, Hugo V and Barnes, Crispin H W},
title = {Quantifying the advantages of applying quantum approximate algorithms to portfolio optimisation},
journal = {Quantum Science and Technology},
}

@article{Sherrington1975,
  title = {Solvable Model of a Spin-Glass},
  volume = {35},
  ISSN = {0031-9007},
  url = {http://dx.doi.org/10.1103/PhysRevLett.35.1792},
  DOI = {10.1103/physrevlett.35.1792},
  number = {26},
  journal = {Physical Review Letters},
  publisher = {American Physical Society (APS)},
  author = {Sherrington,  David and Kirkpatrick,  Scott},
  year = {1975},
  month = dec,
  pages = {1792–1796}
}

@article{Parisi1980,
  title = {A sequence of approximated solutions to the S-K model for spin glasses},
  volume = {13},
  ISSN = {1361-6447},
  url = {http://dx.doi.org/10.1088/0305-4470/13/4/009},
  DOI = {10.1088/0305-4470/13/4/009},
  number = {4},
  journal = {Journal of Physics A: Mathematical and General},
  publisher = {IOP Publishing},
  author = {Parisi,  G},
  year = {1980},
  month = apr,
  pages = {L115–L121}
}

@article{Mezard1986,
  title = {SK Model: The Replica Solution without Replicas},
  volume = {1},
  ISSN = {1286-4854},
  url = {http://dx.doi.org/10.1209/0295-5075/1/2/006},
  DOI = {10.1209/0295-5075/1/2/006},
  number = {2},
  journal = {Europhysics Letters (EPL)},
  publisher = {IOP Publishing},
  author = {Mézard,  M and Parisi,  G and Virasoro,  M. A},
  year = {1986},
  month = jan,
  pages = {77–82}
}

@article{Barahona1982,
  title = {On the computational complexity of Ising spin glass models},
  volume = {15},
  ISSN = {1361-6447},
  url = {http://dx.doi.org/10.1088/0305-4470/15/10/028},
  DOI = {10.1088/0305-4470/15/10/028},
  number = {10},
  journal = {Journal of Physics A: Mathematical and General},
  publisher = {IOP Publishing},
  author = {Barahona,  F},
  year = {1982},
  month = oct,
  pages = {3241–3253}
}

@misc{Apte2022,
      title={Non-Convex Optimization by Hamiltonian Alternation}, 
      author={Anuj Apte and Kunal Marwaha and Arvind Murugan},
      year={2022},
      eprint={2206.14072},
      archivePrefix={arXiv},
      primaryClass={cond-mat.dis-nn},
      url={https://arxiv.org/abs/2206.14072}, 
}

@article{he2025non,
  title={Non-Variational Quantum Random Access Optimization with Alternating Operator Ansatz},
  author={He, Zichang and Raymond, Rudy and Shaydulin, Ruslan and Pistoia, Marco},
  journal={arXiv preprint arXiv:2502.04277},
  year={2025}
}

@article{hao2024end,
  title={End-to-end protocol for high-quality {QAOA} parameters with few shots},
  author={Hao, Tianyi and He, Zichang and Shaydulin, Ruslan and Larson, Jeffrey and Pistoia, Marco},
  journal={arXiv preprint arXiv:2408.00557},
  year={2024}
}

@inproceedings{Basso2022,
  doi = {10.4230/LIPICS.TQC.2022.7},
  url = {https://drops.dagstuhl.de/entities/document/10.4230/LIPIcs.TQC.2022.7},
  author = {Basso,  Joao and Farhi,  Edward and Marwaha,  Kunal and Villalonga,  Benjamin and Zhou,  Leo},
  language = {en},
  title = {The Quantum Approximate Optimization Algorithm at High Depth for MaxCut on Large-Girth Regular Graphs and the Sherrington-Kirkpatrick Model},
  publisher = {Schloss Dagstuhl – Leibniz-Zentrum f\"{u}r Informatik},
  year = {2022},
  copyright = {Creative Commons Attribution 4.0 International license}
}

@article{Farhi2022,
  title = {The Quantum Approximate Optimization Algorithm and the Sherrington-Kirkpatrick Model at Infinite Size},
  volume = {6},
  ISSN = {2521-327X},
  url = {http://dx.doi.org/10.22331/q-2022-07-07-759},
  DOI = {10.22331/q-2022-07-07-759},
  journal = {Quantum},
  publisher = {Verein zur Forderung des Open Access Publizierens in den Quantenwissenschaften},
  author = {Farhi,  Edward and Goldstone,  Jeffrey and Gutmann,  Sam and Zhou,  Leo},
  year = {2022},
  month = jul,
  pages = {759}
}

@article{Venturelli2015,
  title = {Quantum Optimization of Fully Connected Spin Glasses},
  volume = {5},
  ISSN = {2160-3308},
  url = {http://dx.doi.org/10.1103/PhysRevX.5.031040},
  DOI = {10.1103/physrevx.5.031040},
  number = {3},
  journal = {Physical Review X},
  publisher = {American Physical Society (APS)},
  author = {Venturelli,  Davide and Mandrà,  Salvatore and Knysh,  Sergey and O’Gorman,  Bryan and Biswas,  Rupak and Smelyanskiy,  Vadim},
  year = {2015},
  month = sep 
}

@article{HibatAllah2021,
  title = {Variational neural annealing},
  volume = {3},
  ISSN = {2522-5839},
  url = {http://dx.doi.org/10.1038/s42256-021-00401-3},
  DOI = {10.1038/s42256-021-00401-3},
  number = {11},
  journal = {Nature Machine Intelligence},
  publisher = {Springer Science and Business Media LLC},
  author = {Hibat-Allah,  Mohamed and Inack,  Estelle M. and Wiersema,  Roeland and Melko,  Roger G. and Carrasquilla,  Juan},
  year = {2021},
  month = oct,
  pages = {952–961}
}

@article{kremenetski2021quantum,
  title={Quantum alternating operator ansatz (QAOA) phase diagrams and applications for quantum chemistry},
  author={Kremenetski, Vladimir and Hogg, Tad and Hadfield, Stuart and Cotton, Stephen J and Tubman, Norm M},
  journal={arXiv preprint arXiv:2108.13056},
  year={2021}
}

@article{kremenetski2023quantum,
  title={Quantum alternating operator ansatz (qaoa) beyond low depth with gradually changing unitaries},
  author={Kremenetski, Vladimir and Apte, Anuj and Hogg, Tad and Hadfield, Stuart and Tubman, Norm M},
  journal={arXiv preprint arXiv:2305.04455},
  year={2023}
}

@ARTICLE{2020SciPy-NMeth,
  author  = {Virtanen, Pauli and Gommers, Ralf and Oliphant, Travis E. and
            Haberland, Matt and Reddy, Tyler and Cournapeau, David and
            Burovski, Evgeni and Peterson, Pearu and Weckesser, Warren and
            Bright, Jonathan and {van der Walt}, St{\'e}fan J. and
            Brett, Matthew and Wilson, Joshua and Millman, K. Jarrod and
            Mayorov, Nikolay and Nelson, Andrew R. J. and Jones, Eric and
            Kern, Robert and Larson, Eric and Carey, C J and
            Polat, {\.I}lhan and Feng, Yu and Moore, Eric W. and
            {VanderPlas}, Jake and Laxalde, Denis and Perktold, Josef and
            Cimrman, Robert and Henriksen, Ian and Quintero, E. A. and
            Harris, Charles R. and Archibald, Anne M. and
            Ribeiro, Ant{\^o}nio H. and Pedregosa, Fabian and
            {van Mulbregt}, Paul and {SciPy 1.0 Contributors}},
  title   = {{{SciPy} 1.0: Fundamental Algorithms for Scientific
            Computing in Python}},
  journal = {Nature Methods},
  year    = {2020},
  volume  = {17},
  pages   = {261--272},
  adsurl  = {https://rdcu.be/b08Wh},
  doi     = {10.1038/s41592-019-0686-2},
}

@article{Cooley1965,
  title = {An algorithm for the machine calculation of complex Fourier series},
  volume = {19},
  ISSN = {1088-6842},
  url = {http://dx.doi.org/10.1090/S0025-5718-1965-0178586-1},
  DOI = {10.1090/s0025-5718-1965-0178586-1},
  number = {90},
  journal = {Mathematics of Computation},
  publisher = {American Mathematical Society (AMS)},
  author = {Cooley,  James W. and Tukey,  John W.},
  year = {1965},
  pages = {297–301}
}

@inproceedings{Shaydulin2019,
  title = {Multistart Methods for Quantum Approximate optimization},
  url = {http://dx.doi.org/10.1109/HPEC.2019.8916288},
  DOI = {10.1109/hpec.2019.8916288},
  booktitle = {2019 IEEE High Performance Extreme Computing Conference (HPEC)},
  publisher = {IEEE},
  author = {Shaydulin,  Ruslan and Safro,  Ilya and Larson,  Jeffrey},
  year = {2019},
  month = sep 
}

@article{He2023,
  title = {Alignment between initial state and mixer improves QAOA performance for constrained optimization},
  volume = {9},
  ISSN = {2056-6387},
  url = {http://dx.doi.org/10.1038/s41534-023-00787-5},
  DOI = {10.1038/s41534-023-00787-5},
  number = {1},
  journal = {npj Quantum Information},
  publisher = {Springer Science and Business Media LLC},
  author = {He,  Zichang and Shaydulin,  Ruslan and Chakrabarti,  Shouvanik and Herman,  Dylan and Li,  Changhao and Sun,  Yue and Pistoia,  Marco},
  year = {2023},
  month = nov 
}

@inproceedings{Hao2022,
  title = {Exploiting In-Constraint Energy in Constrained Variational Quantum Optimization},
  url = {http://dx.doi.org/10.1109/qcs56647.2022.00017},
  DOI = {10.1109/qcs56647.2022.00017},
  booktitle = {2022 IEEE/ACM Third International Workshop on Quantum Computing Software (QCS)},
  publisher = {IEEE},
  author = {Hao,  Tianyi and Shaydulin,  Ruslan and Pistoia,  Marco and Larson,  Jeffrey},
  year = {2022},
  month = nov,
  pages = {100–106}
}

@article{Shaydulin2021,
  title = {Classical symmetries and the Quantum Approximate Optimization Algorithm},
  volume = {20},
  ISSN = {1573-1332},
  url = {http://dx.doi.org/10.1007/s11128-021-03298-4},
  DOI = {10.1007/s11128-021-03298-4},
  number = {11},
  journal = {Quantum Information Processing},
  publisher = {Springer Science and Business Media LLC},
  author = {Shaydulin,  Ruslan and Hadfield,  Stuart and Hogg,  Tad and Safro,  Ilya},
  year = {2021},
  month = oct 
}

@article{ebadi:22,
	doi = {10.1126/science.abo6587},
	url = {https://doi.org/10.1126/science.abo6587},
	year = 2022,
	month = {jun},
	publisher = {American Association for the Advancement of Science ({AAAS})},
	volume = {376},
	number = {6598},
	pages = {1209--1215},
	author = {S. Ebadi and A. Keesling and M. Cain and T. T. Wang and H. Levine and D. Bluvstein and G. Semeghini and A. Omran and J.-G. Liu and R. Samajdar and X.-Z. Luo and B. Nash and X. Gao and B. Barak and E. Farhi and S. Sachdev and N. Gemelke and L. Zhou and S. Choi and H. Pichler and S.-T. Wang and M. Greiner and V. Vuleti{\'{c}
} and M. D. Lukin},
	title = {{Quantum optimization of Maximum Independent Set using Rydberg atom arrays}},
	journal = {Science}
}

@article{andrist:23,
  title = {Hardness of the maximum-independent-set problem on unit-disk graphs and prospects for quantum speedups},
  author = {Andrist, Ruben S. and Schuetz, Martin J. A. and Minssen, Pierre and Yalovetzky, Romina and Chakrabarti, Shouvanik and Herman, Dylan and Kumar, Niraj and Salton, Grant and Shaydulin, Ruslan and Sun, Yue and Pistoia, Marco and Katzgraber, Helmut G.},
  journal = {Phys. Rev. Res.},
  volume = {5},
  issue = {4},
  pages = {043277},
  numpages = {12},
  year = {2023},
  month = {Dec},
  publisher = {American Physical Society},
  doi = {10.1103/PhysRevResearch.5.043277},
  url = {https://link.aps.org/doi/10.1103/PhysRevResearch.5.043277}
}

@book{Gottlieb1977,
  title = {Numerical Analysis of Spectral Methods: Theory and Applications},
  ISBN = {9781611970425},
  url = {http://dx.doi.org/10.1137/1.9781611970425},
  DOI = {10.1137/1.9781611970425},
  publisher = {Society for Industrial and Applied Mathematics},
  author = {Gottlieb,  David and Orszag,  Steven A.},
  year = {1977},
  month = jan 
}

@BOOK{Boyd2001,
  title     = "Chebyshev and Fourier Spectral Methods",
  author    = "{John P. Boyd}",
  publisher = "Dover Publications",
  series    = "Dover Books on Mathematics",
  edition   =  2,
  month     =  dec,
  year      =  2001,
  address   = "Mineola, NY"
}

@book{Trefethen2000,
  title = {Spectral Methods in MATLAB},
  ISBN = {9780898719598},
  url = {http://dx.doi.org/10.1137/1.9780898719598},
  DOI = {10.1137/1.9780898719598},
  publisher = {Society for Industrial and Applied Mathematics},
  author = {Trefethen,  Lloyd N.},
  year = {2000},
  month = jan 
}

@inproceedings{Lykov2023,
  series = {SC-W 2023},
  title = {Fast Simulation of High-Depth {QAOA} Circuits},
  url = {http://dx.doi.org/10.1145/3624062.3624216},
  DOI = {10.1145/3624062.3624216},
  booktitle = {Proceedings of the SC ’23 Workshops of The International Conference on High Performance Computing,  Network,  Storage,  and Analysis},
  publisher = {ACM},
  author = {Lykov,  Danylo and Shaydulin,  Ruslan and Sun,  Yue and Alexeev,  Yuri and Pistoia,  Marco},
  year = {2023},
  month = nov,
  collection = {SC-W 2023}
}

@misc{boulebnane2025,
      title={Evidence that the Quantum Approximate Optimization Algorithm Optimizes the Sherrington-Kirkpatrick Model Efficiently in the Average Case}, 
      author={Sami Boulebnane and Abid Khan and Minzhao Liu and Jeffrey Larson and Dylan Herman and Ruslan Shaydulin and Marco Pistoia},
      year={2025},
      eprint={2505.07929},
      archivePrefix={arXiv},
      primaryClass={quant-ph},
      url={https://arxiv.org/abs/2505.07929}, 
}

@article{singer2009nelder,
  title={Nelder-mead algorithm},
  author={Singer, Sa{\v{s}}a and Nelder, John},
  journal={Scholarpedia},
  volume={4},
  number={7},
  pages={2928},
  year={2009}
}

@book{conn2009introduction,
  title={Introduction to derivative-free optimization},
  author={Conn, Andrew R and Scheinberg, Katya and Vicente, Luis N},
  year={2009},
  publisher={SIAM}
}

@article{powell2009bobyqa,
  title={The BOBYQA algorithm for bound constrained optimization without derivatives},
  author={Powell, Michael JD and others},
  journal={Cambridge NA Report NA2009/06, University of Cambridge, Cambridge},
  volume={26},
  pages={26--46},
  year={2009}
}

@BOOK{Cheney1998-pn,
  title     = "Introduction to Approximation Theory",
  author    = "Cheney, E W",
  publisher = "American Mathematical Society",
  series    = "AMS Chelsea Publishing",
  edition   =  2,
  month     =  oct,
  year      =  1998,
  address   = "Providence, RI"
}

@BOOK{Jackson2005-ir,
  title     = "The theory of approximation",
  author    = "Jackson, Dunham",
  publisher = "American Mathematical Society",
  series    = "American Mathematical Society Colloquium Publications",
  month     =  may,
  year      =  2005,
  address   = "Providence, RI"
}

@book{Conn2009,
  title = {Introduction to Derivative-Free Optimization},
  ISBN = {9780898718768},
  url = {http://dx.doi.org/10.1137/1.9780898718768},
  DOI = {10.1137/1.9780898718768},
  publisher = {Society for Industrial and Applied Mathematics},
  author = {Conn,  Andrew R. and Scheinberg,  Katya and Vicente,  Luis N.},
  year = {2009},
  month = jan 
}

@article{Sweke2020,
  title = {Stochastic gradient descent for hybrid quantum-classical optimization},
  volume = {4},
  ISSN = {2521-327X},
  url = {http://dx.doi.org/10.22331/q-2020-08-31-314},
  DOI = {10.22331/q-2020-08-31-314},
  journal = {Quantum},
  publisher = {Verein zur Forderung des Open Access Publizierens in den Quantenwissenschaften},
  author = {Sweke,  Ryan and Wilde,  Frederik and Meyer,  Johannes and Schuld,  Maria and Faehrmann,  Paul K. and Meynard-Piganeau,  Barthélémy and Eisert,  Jens},
  year = {2020},
  month = aug,
  pages = {314}
}

\appendix


\section{Additional results on scaling of QAOA depth sufficient to solve SK model exactly}\label{sec:appendix_extra_SK_numerics}

\begin{figure}[h]
    \centering
    \includegraphics[width=0.6\textwidth]{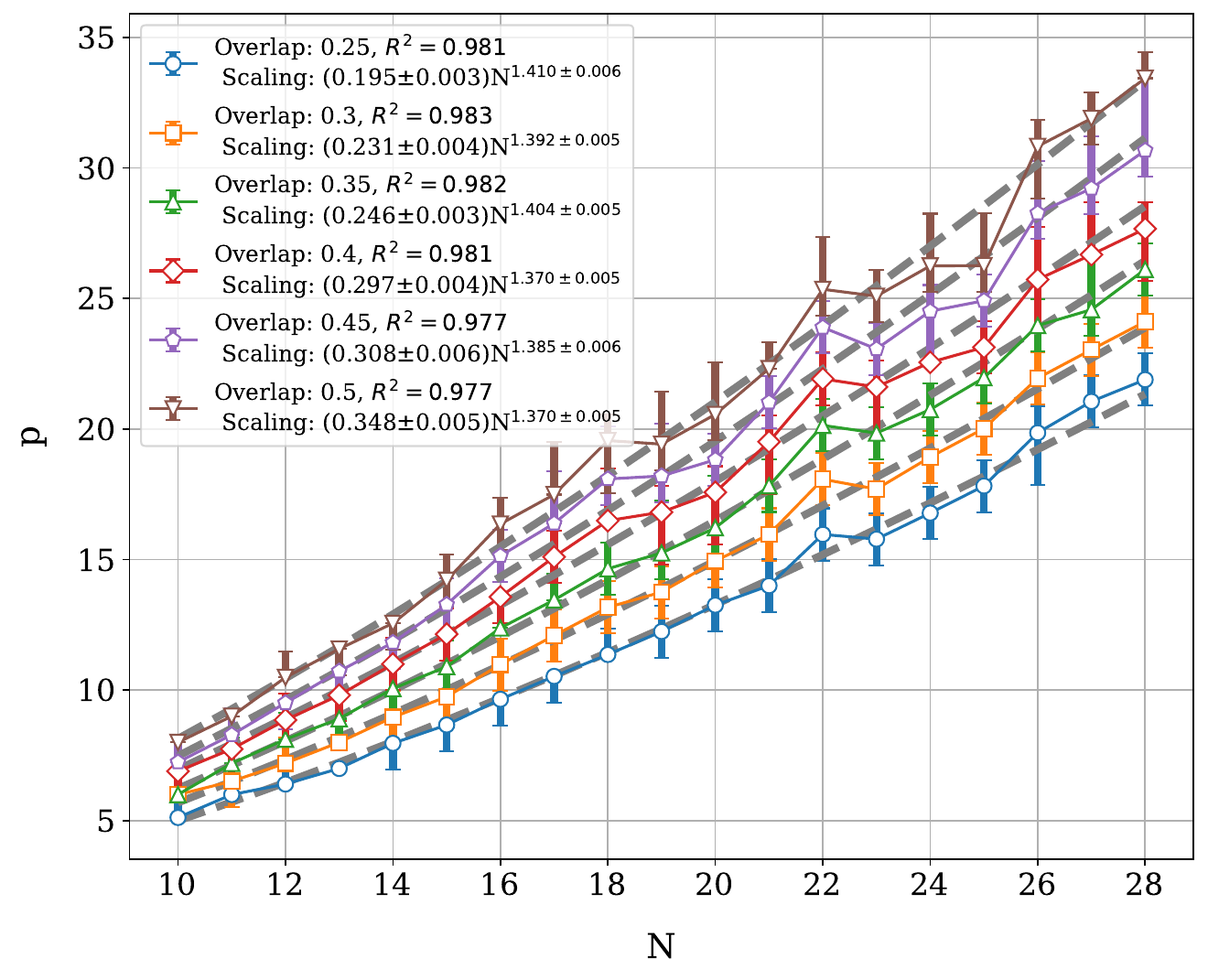}
    \caption{\label{fig:sk_scaling_all_overlaps}
    The scaling of QAOA circuit depth as a function of N for the SK model with different percentile overlaps with the exact ground state.}
\end{figure}

In the main text, we presented the scaling of QAOA circuit depth required to reach a $25\%$ overlap with the ground state for the SK model. Here, we extend our analysis to different overlap thresholds to provide a more comprehensive picture of how QAOA performance scales with problem size. Figure~\ref{fig:sk_scaling_all_overlaps} shows the circuit depth required to achieve overlaps of 25\%, 30\%, 35\%, 40\%, 45\%, and 50\% with the exact ground state as a function of system size $N$. The errors bars correspond to $5 \%$ of points on either side of the median. As expected, higher overlap requirements necessitate deeper circuits, with the 50\% overlap requirement leading to highest $p$ values.

\begin{figure}[htp]
    \centering
    \includegraphics[width=0.6\textwidth]{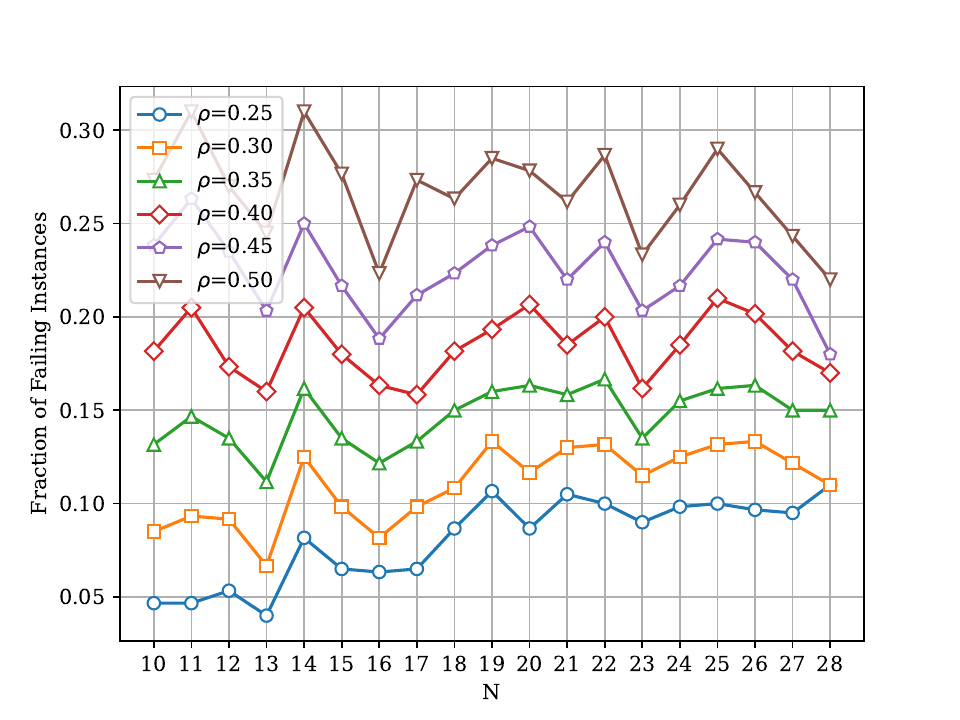}
    \caption{\label{fig:sk_failing_instances}
    The fraction of failing instances with varying overlaps with the exact ground state at each $N$. The failure here corresponds to the run not converging to the desired overlap $\rho$ within the number of evaluations of 40000 or $p_{max} = 2000$.}
\end{figure}

It is important to note that not all instances were able to reach the target overlaps within our constraint. The convergence criteria is getting to the desired overlap within a predefined limit on the number of evaluations (40000) or a maximum depth of $p_{max} = 2000$. For any given value of $N$, there are sets of couplings $J_{ij}$ which lead to much harder instances that require significantly deeper circuits to solve.

Figure~\ref{fig:sk_failing_instances} shows the fraction of instances that failed to reach the specified overlap thresholds for each system size. The failure rate increases with both system size and target overlap, suggesting that there may be fundamental limitations to the ability of QAOA to solve the hardest instances efficiently, that even deeper circuits would be required, or that optimal parameters were not found by our method.

For the instances that did reach the target overlaps, we fit polynomial scaling models as described in Section V.B of the main text. Table~\ref{tab:polynomial_fits} presents the coefficients and goodness-of-fit metrics for each overlap threshold. We also examined exponential fit models (of the form $p = a \cdot b^N$) for comparison, but found consistently better goodness-of-fit metrics for the polynomial models across all overlap thresholds. While we observe good fit quality across all overlap thresholds (with $R^2 > 0.95$), we emphasize that these results should be interpreted cautiously given the limited range of system sizes. The observed exponents $b$ are slightly different for each value of the overlap, but nonetheless are in agreement with each other.

\begin{table}[ht]
\centering
\begin{tabular}{|c|c|c|c|}
\hline
\textbf{Overlap} & \textbf{$a$ coefficient} & \textbf{$b$ exponent} & \textbf{$R^2$} \\
\hline
25\% & 0.19 & 1.41 & 0.981 \\
\hline
30\% & 0.23 & 1.39 & 0.983 \\
\hline
35\% & 0.25 & 1.40 & 0.982 \\
\hline
40\% & 0.30 & 1.37 & 0.981 \\
\hline
45\% & 0.31 & 1.38 & 0.977 \\
\hline
50\% & 0.35 & 1.37 & 0.977 \\
\hline
\end{tabular}
\caption{\label{tab:polynomial_fits}Polynomial fit parameters for the scaling of QAOA depth $p = a \cdot N^b$ required to achieve specified overlaps with the ground state for the SK model.}
\end{table}

Our results suggest that while the required QAOA depth does grow with system size, the growth is polynomial in the average case for the SK model. This is in stark contrast to the LABS problem, where the depth requirement appears to grow exponentially. 

\section*{Disclaimer}
This paper was prepared for informational purposes by the Global Technology Applied Research center of JPMorganChase. This paper is not a product of the Research Department of JPMorganChase or its affiliates. Neither JPMorganChase nor any of its affiliates makes any explicit or implied representation or warranty and none of them accept any liability in connection with this position paper, including, without limitation, with respect to the completeness, accuracy, or reliability of the information contained herein and the potential legal, compliance, tax, or accounting effects thereof. This document is not intended as investment research or investment advice, or as a recommendation, offer, or solicitation for the purchase or sale of any security, financial instrument, financial product or service, or to be used in any way for evaluating the merits of participating in any transaction.

\end{document}